\documentclass[jcp,amsmath,amssymb,showkeys,floatfix,
reprint
]{revtex4-1}

\usepackage{bibentry}
\usepackage{appendix}
\usepackage[framemethod=tikz]{mdframed}
\mdfsetup{
	outerlinewidth=1pt,
	linecolor=white!90!black,
	innertopmargin=6pt,
	innerbottommargin=6pt,
	leftmargin=2pt,
	rightmargin=2pt
}
\usepackage[utf8]{inputenc}
\usepackage{float}
\usepackage{fixltx2e}
\usepackage{graphicx}

\usepackage{color}

\usepackage{graphicx}
\usepackage{dcolumn}
\usepackage{bm}
\usepackage{xfrac}
\usepackage{colortbl}

\usepackage{subfigure}

\usepackage{natbib}
\usepackage{bibentry}
\usepackage{float}
\usepackage{sidecap}
\usepackage[section]{placeins}

\usepackage[caption=false]{subfig}

\usepackage{amsmath}
\usepackage{amssymb}
\usepackage{physics}
\usepackage{marvosym}
\usepackage{wasysym}
\usepackage{pifont}
\usepackage{wrapfig}
\usepackage{xifthen}
\usepackage{enumerate}
\usepackage{cancel}
\usepackage{MnSymbol}
\usepackage{mathtools}
\usepackage{relsize}

\newcommand{\Fig}[1]{Fig.~\ref{f:#1}}

\newcommand{\Eq}[1]{Eq.~\eqref{e:#1}}

\newcommand{\Section}[1]{Section~\ref{s:#1}}

\usepackage[normalem]{ulem}

\begin{document}
\title{Simulation of adiabatic quantum computing for molecular ground states}

\author{Vladimir Kremenetski }
\affiliation{Quantum Artificial Intelligence Laboratory (QuAIL), Exploration Technology Directorate, NASA Ames Research Center, Moffett Field, CA 94035, USA}
\affiliation{USRA Research Institute for Advanced Computer Science, Mountain View, California 94043, USA}

\author{ Carlos Mejuto-Zaera}
\affiliation{Computational Research Division, Lawrence Berkeley National Laboratory, Berkeley, California 94720, USA}
\affiliation{Department of Chemistry, University of California, Berkeley, California 94720, USA}

\author{Stephen J. Cotton}
\affiliation{Quantum Artificial Intelligence Laboratory (QuAIL), Exploration Technology Directorate, NASA Ames Research Center, Moffett Field, CA 94035, USA}
\affiliation{KBR, 601 Jefferson St., Houston, TX 77002}

\author{ Norm M. Tubman}
\email{norman.m.tubman@nasa.gov}
\affiliation{Quantum Artificial Intelligence Laboratory (QuAIL), Exploration Technology Directorate, NASA Ames Research Center, Moffett Field, CA 94035, USA}

\date{\today}
\begin{abstract}
	Quantum computation promises to provide substantial speedups in many practical applications with a particularly exciting one being the simulation of quantum many-body systems. Adiabatic state preparation (ASP) is one way that quantum computers could recreate and simulate the ground state of a physical system. In this paper we explore a novel approach for classically simulating the time dynamics of ASP with high accuracy, and with only modest computational resources via an adaptive sampling configuration interaction (ASCI) scheme for truncating the Hilbert space to only the most important determinants. We verify that this truncation introduces negligible error, and use this new approach to simulate ASP for sets of small molecular systems and Hubbard models. Further, we examine two approaches to speeding up ASP when performed on quantum hardware: (i) using the complete active space configuration interaction (CASCI) wavefunction instead of the Hartree-Fock initial state and (ii)~a non-linear interpolation between initial and target Hamiltonians. We find that starting with a CASCI wavefunction with a limited active space yields substantial speedups  for many of the systems examined while non-linear interpolation does not. Additionally, we observe interesting trends in the minimum gap location (based on the initial state) as well as how critical time can depend on certain molecular properties such as the number of valence electrons. Importantly, we find that the required state preparation times do not show an immediate exponential wall that would preclude an efficient run of ASP on actual hardware. 
\end{abstract}

\maketitle

\section{Introduction}
\label{s:intro}
Recent advances in quantum computers have shown much potential to realize new applications beyond the current reach of classical computing~\cite{omalley2016,google2019}. One of the most promising such applications is the simulation of physical  systems that are 'natively' quantum. Of particular interest are simulations in the field of quantum chemistry and material science~\cite{zalka1998}. Many promising algorithms for quantum simulation have been proposed and developed, which include state preparation algorithms for efficiently simulating molecular ground states and calculating their energies ~\cite{bravyi2020}.  


The main approach  of finding ground states for quantum Hamiltonians in the NISQ era is the variational quantum eigensolver (VQE) method. This is a hybrid quantum-classical algorithm in which a ground state candidate  is prepared with a parameterized quantum circuit with variational parameters that can be optimized to minimize the expectation energy. By splitting up the algorithm into a classical optimization and simple quantum energy measurement, the requirements for gate coherence time are much lower~\cite{grimsley2019}. The optimization over the gate parameters can be performed with classical algorithms~\cite{mcclean_2016}. Various optimization methods include approaches with and without gradients~\cite{grimsley2019} which have shown promise with limited testing, but also come with many well known issues~\cite{quantummontecarloreview}.  Some of the problems include optimization algorithms getting stuck in local minima, hitting plateaus in the energy landscape~\cite{mcclean2018}, or having prohibitively high runtime costs that scale with the number of parameters~\cite{wierichs2020, izmaylov2020}. 

Adiabatic state preparation (ASP) is an alternative approach to preparing ground states to study their wavefunction and energy, one which avoids dealing with a noisy optimization problem. This involves starting with a Hamiltonian in which the ground state is known and can be easily prepared, after which the system is time evolved by slowly changing the Hamiltonian into the one whose ground state is required. By the quantum adiabatic theorem, if the Hamiltonian is changed slowly enough, the system will remain in the ground state over the entire evolution thereby reaching  the desired ground state at the end~\cite{hauke_2020}. This approach avoids the complications of optimizing variational parameters in the presence of noise.  

In spite of its clear implications for the many-body problem, to the best of our knowledge only one study of time resource estimates has been performed for implementing ASP with regards to molecular  and material science applications~\cite{veis2014}, with most studies focusing on efficient embedding onto available hardware ~\cite{babbush2014AdiabaticQS, moll2016}.   One reason for this is that the coherence times of current gate model hardware are not well suited for this approach~\cite{hauke_2020}, while VQE works better with limited coherence time on NISQ gate model devices~\cite{grimsley2019}.  
Hardware specific to annealing has also not yet been used for these sorts of simulations because physical Hamiltonians are in general non-stoquastic~\cite{babbush2015,hauke_2020} (that is, having non-negative off-diagonal elements in the computational basis) , however current advances with non-stoquastic couplers suggest that this might change in the near term.  

Recent advances in the implementation of non-stoquastic Hamiltonians on superconducting flux qubits \cite{ozfidan2020} have made research into annealing techniques for quantum simulation  promising. Other advances have also been made to make annealing more practically useful, such as research showing that pausing the evolution just after the minimal gap   during the annealing process makes the final overlap with the desired ground state orders of magnitude higher due to thermalization \cite{marshall2019}. The time window for making this pause can also be regulated through the coupling strength between physical qubits representing a single logical one \cite{gonzales2020}. Though ASP, unlike annealing in general, does not take place at finite temperatures and requires the system to stay in the ground state during the evolution, annealing hardware is still useful for instantiating ASP as these additional restrictions are imposed. This is important to note since, for this paper, we will be focusing solely on ASP.

Given the difficulties of noisy optimization required for VQE and recent advances in annealing hardware, it is important to consider how well state preparation can be performed with ASP when applied to quantum simulations.   Although much has been done on annealing hardware with classical optimization problems, only a small number of studies have been performed with regards to quantum simulation.   One of the few detailed works in this regard looked at molecular systems~\cite{veis2014} and considered several approaches to improve the adiabatic process through a case study of methylene. Their study  included testing the effect of different starting Hamiltonians, doing a non-linear time evolution between the initial and final Hamiltonians, and using different quality starting states such as those from a small complete active space configuration interaction (CASCI) wavefunctions.

Against this backdrop, the purpose of this paper is to follow up on some of the molecular results described in \cite{veis2014} by applying, for the first time, a time-dynamics approach using a selective CI protocol to verify their results for a much wider array of different physical systems. Using selective CI-based dynamics is essential, since it allows us to expand the complexity of the Hamiltonians considered thanks to the ability to include a much larger number of spatial orbitals.
~The relationship between the initial overlap of the starting Hamiltonian's ground state with the target ground state and the time required for state preparation is examined across fourteen different molecules, as well as a handful of Hubbard model instances at half-filling. Our proxy for the state preparation time here is the minimum amount of time required to achieve a squared overlap with the target state close to 1. The reason for focusing on final squared overlap as a criterion for a concluded simulation lies its relevance in the phase estimation algorithm, where the squared overlap with the ground state corresponds to the probability of observing the ground state energy \cite{aspuru2005}.

Using this approach, we find that the use of CASCI wavefunctions as initial ground states results in a speedup of a factor of 2-4 over using the Hartree-Fock initial ground state, with non-linear interpolation between initial and target Hamiltonians producing negligible results (though likely due to minor changes in the energy gap during the evolution).
In the next section, we will give an overview of ASP principles and theory, a brief review of Lanczos dynamics and a more in-depth discussion of ASCI and theoretical motivation for this paper. Afterwards, we give a detailed description of our results followed by a discussion on useful takeaways from the research and potential focuses for future projects. 

\section{Theoretical Background and Methods}

\subsection{Principles of adiabatic evolution and state preparation}
\newcommand{\nth}{$n^{\text{th}}$}

ASP works by leveraging the Quantum Adiabatic Theorem, which states that if a system is in the \nth~eigenstate of a Hamiltonian, then if that Hamiltonian is changed sufficiently slowly (and there are no spectral crossings with the \nth~energy) the system will still be in the \nth~state of the resulting Hamiltonian at the end. Consequently, ASP exploits this by first preparing the qubits in the ground state of a Hamiltonian ($H_{\text{initial}}$) whose ground state is easy to prepare directly, and then slowly changing the Hamiltonian to the one, $H_{\text{target}}$, whose ground state is needed, over a period of time $T$. This change of Hamiltonians is done most simply through a direct linear interpolation between them:
\begin{equation}
H(s) = (1-s)H_{\text{initial}} + sH_{\text{target}},\ 0 \leq s \leq 1,\ s=\frac{t}{T} 
\label{eq:1}
\end{equation}
Though ASP is not more computationally powerful than other quantum computing methods~\cite{aharonov2008}, ASP can provide practical speedups over conventional state preparation methods. However, a key limitation that it faces is the amount of time $T$ that it spends evolving the Hamiltonian. Through another result of the quantum adiabatic theorem,

\begin{equation}
	T >> \frac{\epsilon}{g_{\text{min}}^2} 
\label{eq:critical_time_constraints}
\end{equation}

where $g_{\text{min}}$ is the minimum energy gap:

\begin{equation}
g_{\text{min}} = \min\limits_{0\leq s\leq 1} [ E_1(s)-E_0(s)] 
\end{equation}

\noindent
and 

\begin{equation}
\epsilon = \max\limits_{0\leq s \leq 1}|\bra{\lambda_1(s)} \frac{d H}{d s}\ket{\lambda_0(s)}| 
\end{equation}
\noindent
where $\ket{\lambda_1(s)} $, $E_1(s)$ are the first excited state and energy, and $\ket{\lambda_0(s)} $, $E_0(s)$ are the ground state and energy of $H(s)$.

Since $\epsilon$ scales at most polynomially with system size~\cite{yung2012}, most difficult problems for ASP stem from a $g_{\text{min}}$ that shrinks exponentially with the size of the system. However, on a practical level, reducing $\epsilon$ by large factors can still bring certain problems into the realm of computational feasibility, and can be used in conjunction with other methods to accelerate the preparation process. One such method of reducing $\epsilon$ was discussed in a paper by Veiss and Pipner~\cite{veis2014}, in which they proposed the approach of choosing an initial ground state $\ket{\psi_0}$ with a relatively high degree of overlap with the target ground state $\ket{\psi_t}$. In many cases, finding such a $\ket{\psi_0}$ can be efficiently done using classical methods for approximating the target state $\ket{\psi_t}$, whose result might not be close enough as a direct chemical simulation of $\ket{\psi_t}$  but is good enough for the purposes of ASP.  

\subsection{Adiabatic quantum dynamics via Lanczos-based time-propagation}
\label{s:Lanczos}
As the ground state is evolved through a Hamiltonian that changes with time, the evolution of the state is governed by the time-dependent Schrodinger Equation: 
\begin{equation}\label{e:TDSE}
i\hbar \frac{\partial}{\partial t}\ket{\Psi(t)} = H(t)\ket{\Psi(t)} 
\end{equation}
For a time-dependent Hamiltonian, the solution can be approximated numerically by breaking up the exponentiated operator across several small time steps
\[ \ket{\Psi(T)} \approx e^{-\frac i\hbar H(T)\Delta t} \, \cdots\, e^{-\frac i\hbar H(\Delta t)\Delta t}\, e^{-\frac i\hbar H(0)\Delta t} \ket{\Psi(0)}.  \]
A way is then needed to efficiently compute the multiplication  $e^{-\frac i\hbar H(t_i)\Delta t} \ket{\Psi(t_i)} $ with exponentiated, potentially large $H(t_i)$. 

In order to calculate this, a 
Lanczos-based time-evolution algorithm is applied~\cite{park1986,manmana2005}, which is commonly used to study correlated dynamics in many-body systems~\cite{kollath2007,carleo2012,pastori2019}. 
Noting that the time-evolution may be represented as power-series in $H$, of low order for sufficiently small~$\Delta t$, a Krylov (power) space is constructed through successive action of $H$ on the starting wavefunction $\ket{\Psi}$; I.e., the usual 3-term Lanczos recursion~\cite{park1986} is leveraged,

\begin{subequations}
\begin{equation}
\ket{\tilde v_{i+1}} = H\ket{v_i} - \bra{v_i}H\ket{v_i}\ket{v_i} - \bra{v_{i-1}}H\ket{v_i}\ket{v_{i-1}} ,
\end{equation}
with normalization
\begin{equation}
\ket{v_{i+1}} = \ket{\tilde v_{i+1}} / \sqrt{\bra{\tilde v_{i+1}}\ket{\tilde v_{i+1}}} ,
\end{equation}
\end{subequations}
using $\ket{\Psi}$ as the seed $\ket{v_0}$ vector. For each element $\ket{v_{i+1}}$ added to the basis, a representation of the Hamiltonian $H(t)$ is constructed in the subspace ($H_{\text{Krylov}}$) where it takes the form of a tridiagonal matrix. This matrix is then exponentiated by finding its eigenvalues $\{\lambda_i\}$ and eigenvectors $\{\ket{\lambda_i}\}$ through direct diagonalization and exponentiating the eigenvalues $\sum_{i=1}^N e^{\frac{i\Delta t}{i\hbar}\lambda_i}\ket{\lambda_i}\bra{\lambda_i}$. Since $\ket{\Psi} = \ket{v_0}$,  the first element of the Krylov basis, $e^{\frac{H_{\text{Krylov}}\Delta t}{i \hbar}}\ket{\Psi}$ is represented by the first row of the exponentiated Krylov matrix, which is all that is calculated. The final element of this row is then checked to see if its magnitude (the contribution by the recently added basis vector to the final result) is below a certain threshold - if not, a new element is added to the Krylov basis and the above process is repeated~\cite{park1986}. Once this iterative process is concluded, the final product is reconstructed from the elements of the Krylov basis. It's important to note that since the evolved state stays pretty close to the instantaneous ground state throughout the adiabatic evolution, the size of the Krylov space required for convergence is usually well under 100, so the direct diagonalization procedure remains computationally manageable.

\subsection{Efficient time-propagation in large orbital basis sets via adaptive sampling configuration interaction}\label{s:ASCI}

Selective CI procedures are routinely used to solve static electronic structure problems. However, they are less frequently employed in the quantum dynamics context, though notable work has recently been done on this front~\cite{guther2018, schriber2019}. In this work, we employ a particular effective form of selective CI known as adaptive sampling configuration interaction (ASCI)~\cite{tubman2016,tubman2018a,tubman2020} to vastly increase the size of the Hilbert space (i.e., the number of 1-particle orbital basis functions) we can accurately simulate with the Lanczos dynamics protocol described in \Section{Lanczos} with almost no loss in accuracy (as verified in \Section{Results}). This enables us, among other things, to use the  cc-pVTZ orbital basis to describe our molecule set, which would require $10^{11}-10^{17}$ determinants if we were to use a full CI procedure. Storing this many determinants, let alone the matrices that would be built using them, would be infeasible even for the most powerful supercomputers, as would the time required to run such a simulation.

Selective CI procedures work by searching and identifying the important determinants in Hilbert space and then truncating the Hilbert space so as to only include these important determinants in subsequent energy optimization/diagonalization steps. 
ASCI identifies the most important determinants by trying to find those determinants for which the ground state wave function has the highest norm coefficients. This process is performed iteratively, with the estimated importance of the CI coefficients of the determinants found in one iteration computed from the CI coefficients of determinants found in a previous iteration using a continuity equation. To arrive at this equation, suppose there was an exact representation of the desired ground state wavefunction using Slater determinants $\{\ket{D_i}\}$ with exact CI coefficients $\{C_i\}$~\cite{tubman2016}:
$$\ket{\lambda_0} = \Sigma_iC_i\ket{D_i} $$
Applying the Hamiltonian H to either side would give:
$$H\ket{\lambda_0} = \sum_{i=0}^N (\Sigma_j C_j\bra{D_i}H\ket{D_j})\ket{D_i}$$
$$E_0\Sigma_iC_i\ket{D_i} = \Sigma_i(\Sigma_{j=1}^N C_j\bra{D_i}H\ket{D_j})\ket{D_i},$$
where $E_0$ is the ground state energy. Therefore, for each individual $\ket{D_i}$, there would be:
$$E_0C_i = \Sigma_j C_j\bra{D_i}H\ket{D_j} $$
Solving for $C_i$, gives the continuity equation:
\begin{equation}
C_i = \frac{\Sigma_{j\neq i} C_j\bra{D_i}H\ket{D_j}}{E_0- \bra{D_i}H\ket{D_i}} = - \frac{\sum_{j \ne i} H_{ij} C_j}{H_{ii}-E_0} \label{e:ASCI}
\end{equation}

At iteration $k$, the initial determinants are the best determinants found in the previous iteration $\{D_i\}^k$ and their coefficients $\{C_i\}^k$ (at the very beginning, the initial set of determinants contains just the Hartree-Fock state with a coefficient of 1). The next step involves finding all of the determinants that can be reached from the determinants in $\{D_i\}^k$ with the largest coefficients $C_i$ through single- and double- excitations of electrons to other orbitals. The number of determinants from which  new determinants are found through excitation is the parameter referred to as the "Core Size"~\cite{tubman2020}, which existing research suggests needs to only be a few percent of the maximum number of determinants we save across iterations, the "Target Size"~\cite{tubman2020}.
The coefficients of these newly found determinants (many of them possibly the same as the old ones) are computed using the consistency equation, the previous determinants $\{D_i\}^k$, and the previous estimate of the ground state energy $E_0^k$:

\begin{equation}
C_i^{k+1} = -\frac{\mathlarger{\sum}_{D_j\in \{D\}^k,j\neq i}C_j^k\bra{D_i}H\ket{D_j}}{\bra{D_i}H\ket{D_i} - E_0^k}
\end{equation}

The newly found determinants are then sorted by the magnitude of their coefficients $C_i^{k+1}$, and if the number of determinants exceeds the target size then the determinants with the largest coefficients are kept. A matrix is built for the Hamiltonian $H$ in this basis of determinants, and the coefficients from its constructed ground state vector are used as the final coefficients for the iteration. The ground state energy $E_0^{k+1}$  of $H$ is found to check for convergence. Once the energy changes below a chosen energy convergence tolerance across iterations, the procedure is concluded~\cite{tubman2016}. 
The determinants found through \Eq{ASCI} thus provide a compact space for representing the wavefunction, and diagonalizing the full Hamiltonian within this subspace has been demonstrated \cite{tubman2016,tubman2018a, tubman2018b,hait2019,mejuto2019, tubman2020,eriksen2020} to give very accurate energies for complicated electronic systems exhibiting strong-correlation and many-body effects. 

The novelty of this paper (within the ASCI framework) is that \Eq{ASCI} is to be applied in the context of fermionic dynamics. The key point which enables the set of determinants found through \Eq{ASCI} to be leveraged in computing the adiabatic evolution described above is that the time-dependent Hamiltonian $H(s)$ takes a similar form over the entire course of the evolution and, in fact, the final Hamiltonian, i.e, $H(s=1)$, is known from the beginning. The full/final Hamiltonian may, therefore, simply be used at the outset to identify the determinants by solving the continuity relation in \Eq{ASCI}, and using the resulting truncated space as the basis upon which to build the Krylov subspace described in \Section{Lanczos}. Additionally, since properly modeling the energy gap is an important for modeling adiabatic evolution, we also find the determinants required to build the first excited states. We do so by seeding ASCI with the second-lowest energy Slater determinant and using $E_1$ rather than $E_0$ in the procedure described above. Such adiabatic ASCI dynamics thereby constitutes a powerful and general tool for simulating adiabatic quantum dynamics as described above and demonstrated below. 


\section{Simulation Details}
\label{s:boring_details}

Fourteen molecules from the G1 set are tested in the cc-pVDZ and cc-pVTZ basis sets this work: LiH, HF, NaCl, LiF, Na\textsubscript{2}, CH\textsubscript{2}, H\textsubscript{2}O, CH\textsubscript{4}, HCl, N\textsubscript{2}, P\textsubscript{2}, Cl\textsubscript{2}, Li\textsubscript{2},  and F\textsubscript{2}. Eight additional molecules are tested only with the  cc-pVTZ set (CO\textsubscript{2}, SiH\textsubscript{4}, NH\textsubscript{3}, ClF, SiH\textsubscript{2}, LiNa, NaH, and H\textsubscript{2}). Experimental data for the equilibrium geometries comes from the following reference~\cite{cccbdb2019}.  Hartree-Fock simulations are performed with Psi4~\cite{psi4}. We find the singlet ground states of all molecules (this is the true ground state for all of them but CH\textsubscript{2}, whose ground state is a triplet). 
For the purposes of benchmarking, we perform molecular simulations using full CI with a limited number of spatial orbitals so that we could reasonably run our simulation given memory and simulation time constraints.  The details for these full CI simulations are summarized in Table~\ref{table1}. For the approximate simulations we run ASCI on the target Hamiltonian and find the most important determinant states to accurately describe the ground state and the first excited state (as described in the methods section). For runs using the cc-pVDZ basis we choose a target size of $10^5$, and a core size of equal value. For runs using the cc-pVTZ basis of spatial orbitals (up to 64 orbitals, due to the current implementation), ASCI is also run on the target Hamiltonian, but this time with a target size of $10^6$ and a core size of $2*10^4$. In all cases for ASCI, the frozen core approximation is used, as well as a time-step size of 0.1 (a.u.). For ASCI, an energy convergence tolerance of $10^{-5}$ Ha was used. 

\begin{table}
\begin{center}
\begin{tabular}{ |c|c|c|c|c|c|c|c|c|c|c|c|c|c|c| } 
 \hline
 \textbf{Molecule} & H\textsubscript{2}O & LiH & F\textsubscript{2} & HF & HCl & Li\textsubscript{2} & LiF\\ 
 \textbf{Orbitals} & 17 & 19 &  14 &  17 & 17  & 26  & 17 \\ 
 \textbf{Frozen Core} & Yes & No & Yes & Yes & Yes & No & Yes \\ 
 \textbf{Step Size (a.u.)} & 0.1 & 0.1 & 0.1 & 0.1 & 0.1 & 1.0 & 0.1\\
 \hline

 \hline
 \textbf{Molecule} & N\textsubscript{2} & Cl\textsubscript{2} & P\textsubscript{2} & NaCl & Na\textsubscript{2} & CH\textsubscript{4} & CH\textsubscript{2}\\ 
 \textbf{Orbitals} & 14 & 14 &  15 &  17 & 26  & 16  & 20 \\ 
 \textbf{Frozen Core} & No & Yes & Yes & Yes & Yes & Yes & Yes \\ 
 \textbf{Step Size (a.u.)} & 0.1 & 0.2 & 0.5 & 1.0 & 0.1 & 0.1 & 0.1\\
 \hline
\end{tabular}
\end{center}
\caption{Benchmark simulation details for the adiabatic evolution in the full determinant space (full CI). The choice of time step size and spatial orbital count was decided by limitations on memory and computing time, (within a range that would not produce significant error, see Appendix for time step error). Orbital basis used is cc-pVDZ. \label{table1}}
\end{table}

The spatial orbitals for the Hubbard models are generated using PySCF's self-consistent field method \cite{PYSCF}. The two main models used have 18 sites with periodic boundary conditions and 20 sites with open boundary conditions, both at half-filling and a constant hopping term of $t=1$. Within these two main models, we introduce variations by changing $U$ and $\epsilon$ in the Hamiltonian:
$$H = - t\Sigma_{(i,j)} a_i^\dag a_{j} + \Sigma_i U n_{i,\uparrow} n_{i,\downarrow} + \Sigma_i((-1)^i\epsilon)(n_{i,\uparrow} +n_{i,\downarrow})$$
 We benchmark ASCI's performance against a 3-by-4 Hubbard model with periodic boundary conditions (and a 1-by-12 model with open boundary conditions) at half-filling with exact diagonalization. As with the molecular set, the ASCI energy convergence tolerance is $10^{-5}$ Ha and the time step size is 0.1 (a.u.) for all of the ASCI simulations. For the benchmarking cases, time step sizes of 0.01 (a.u.) are used where the critical time is less than or around 1.0 (a.u.). For the non-benchmarking cases, a target size of $10^6$ and core size of $2*10^4$ were used. 

The starting Hamiltonian used is diagonal in the space of Slater determinants, and shares its diagonal with the target Hamiltonian. This guarantees that the Hartree-Fock state is the ground state. Both this and the target Hamiltonian are built in their matrix representation using either the full space of determinants, or those that are identified by ASCI as being the most important for the construction of the ground or first excited states in the final Hamiltonian. The ground state of the target Hamiltonian is then found using the Lanczos iterative method, and the magnitude of its squared inner product with the starting ground state is found. The latter constitutes the "initial overlap" discussed throughout the paper. The transition from the starting to the target Hamiltonian is done through a simple linear interpolation between the two, as seen in Eq.~\eqref{eq:1}. The adiabatic evolution of the state itself is carried out using Lanczos dynamics. . 

The minimum evolution time required to achieve a final squared overlap of at least 0.99 with the target ground state is found up to two significant figures. Since all of the molecular ground states already have a high degree of overlap with the target ground state, the high cutoff of 0.99 is simply used as a proxy for how well the target ground state was prepared (corresponding to a 99$\%$ chance of the evolved state collapsing to the true ground state in a quantum phase estimation run). For the case of the cc-pVDZ basis we also find the minimum evolution time required for the expectation energy of the prepared state to be within chemical accuracy of the true ground state energy. 
 
Our matrices are represented as lists of 3-tuples containing the two indices and entry of each matrix element, and are stored in CSC format. We implemented our own version of the Lanczos method, with a convergence tolerance of $10^{-12}$ Ha and the vectors taken from the C++ standard library. Direct diagonalization methods were used from the GNU Scientific Library. Slater determinants were represented by C++ bitsets. 
\section{Results}
\label{s:Results}
\subsection{Benchmarking ASCI dynamics for ASP}
\subsubsection{Molecular Set}
\begin{figure}[H]
\centering
\includegraphics[width=\linewidth]{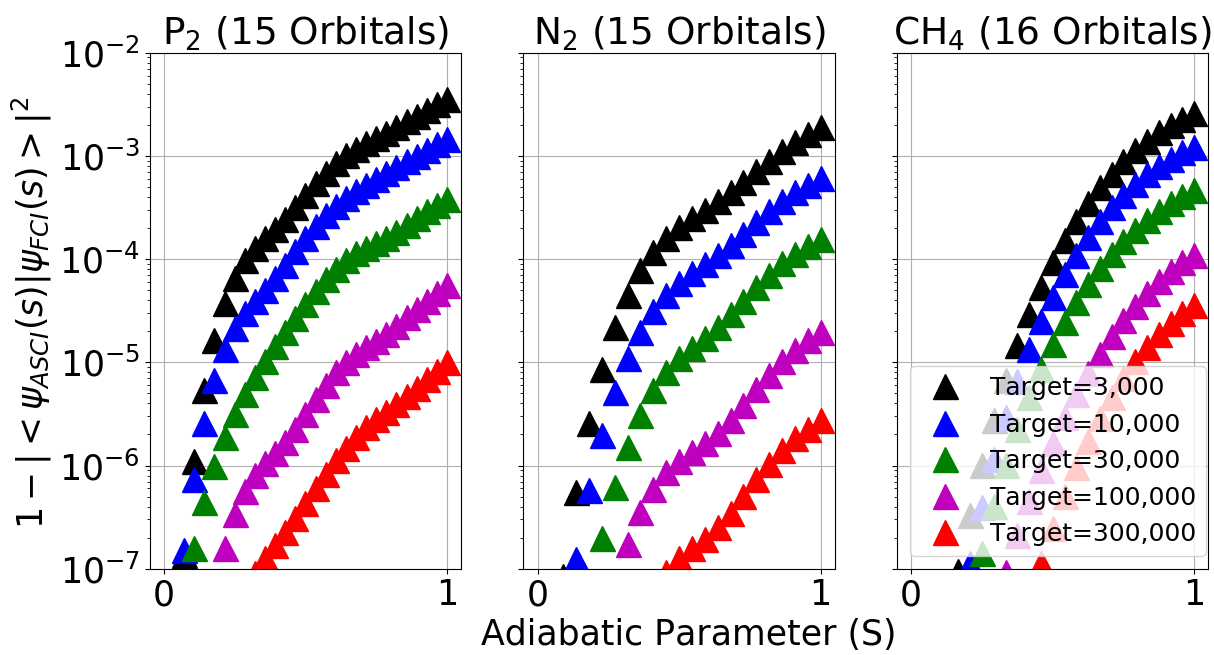}
\caption{ASCI performance in accurately recreating evolved wavefunction for various target sizes (denoted by "target") with cc-pVDZ orbital basis. The error at each time step represents one minus the inner product between the wavefunction constructed using ASCI states and the wavefunction constructed using FCI. The core size is $2\%$ of target size. }
\label{f:asci_evolved_state_error}
\end{figure}

A threshold question addressed by this work is the extent to which the adiabatic ASCI dynamics protocol (as described in \Section{ASCI}) provides an accurate description of the exact fermionic dynamics in this particular application. Establishing this to be the case justifies this work as well as follow-up work seeking to assess ASP's feasibility as a viable quantum computing method for other, perhaps more complicated, atomic, molecular, and solid-state systems. 
As a first diagnostic, we find the squared overlap between the ASCI evolved state and the evolved state constructed using FCI at each time step. We then plot the error from 1 ($1-|\bra{\psi_{\text{ASCI}}(s)}\ket{\psi_{\text{FCI}}(s)}|^2$) as displayed in \Fig{asci_evolved_state_error} for three molecules. The accuracy of ASCI dynamics in describing the evolved state was the same or better for the other molecules that we examined this way. For the systems we examined, $~10^5$ target states were sufficient to achieve at least a 99.99$\%$ overlap with the exact evolved wavefunction at all time steps.

\begin{figure}[H]
\centering
\includegraphics[width=\linewidth]{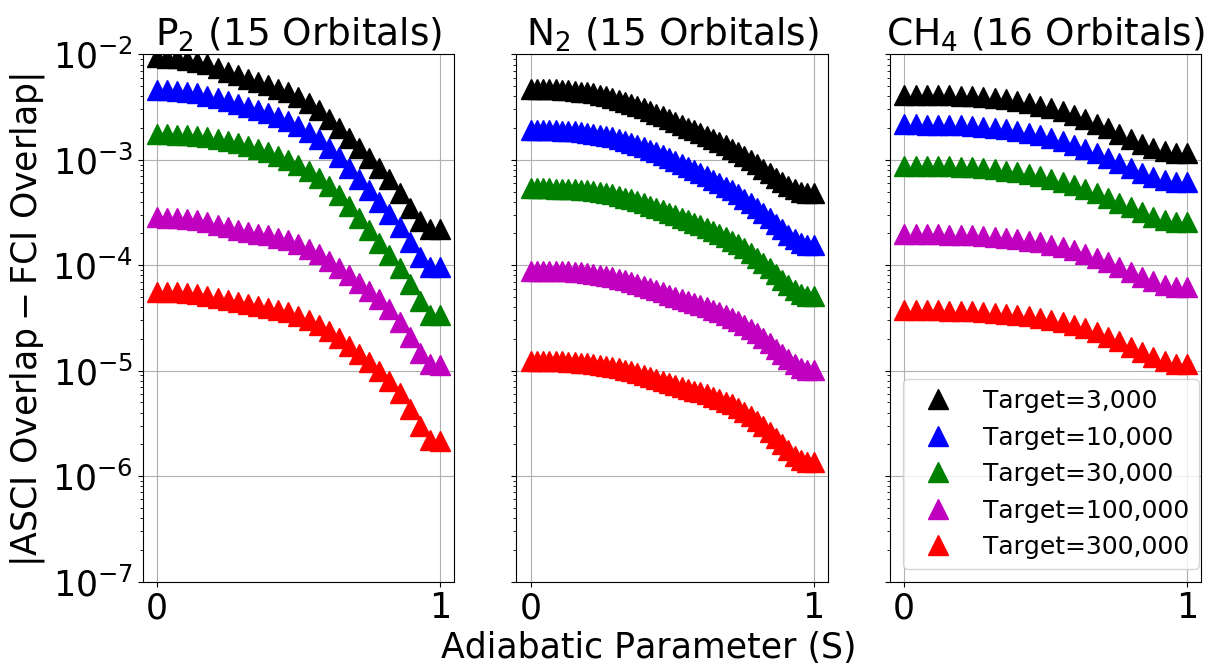}
\caption{ASCI performance for various target sizes (denoted by "target") using cc-pVDZ orbital basis. The error at each time step represents the difference between the evolved state's squared overlap with the target state  predicted using ASCI states and the overlap predicted in the full CI case. The core size is $2\%$ of target size. The error decreases over time since in all cases the squared overlaps are converging to a common limit of 1.}
\label{f:asci_ts_error}
\end{figure}

To be sure that this error in the evolved state does not meaningfully affect our results, we also find the difference in the predicted squared overlap of the prepared and target states between the ASCI and full CI cases at each time step, given by $|\ \bra{\psi_{(s)}^{ASCI}}\ket{\psi_{target}^{ASCI}}^2-\bra{\psi_{(s)}^{FCI}}\ket{\psi_{target}^{FCI}}^2\ |$. This is done for several molecules, the results for three of which are displayed in \Fig{asci_ts_error}.
For these cases, $10^5-10^6$ determinants  are enough to give a negligible worst-case error of $\sim 10^{-4}-10^{-5}$ in the predicted squared overlap of the evolved ground state with target ground state, though for our purposes an error of up to $\sim 5 *  10^{-2}$ would be sufficient to justify the results in this work. 
In addition to checking the accuracy of ASCI for the size of the overlap, we also  calculate the energy gap $E_1(s)-E_0(s)$ over the course of the evolution. The estimate of the gap size is off by at most 6$\%$ for the largest target sizes, which is negligible compared to the change undergone by the energy gap during the evolution and is sufficient for purpose of studying simulated adiabatic evolution. 
\subsubsection{Hubbard Set}

\begin{figure}[H]
\centering
\includegraphics[width=\linewidth]{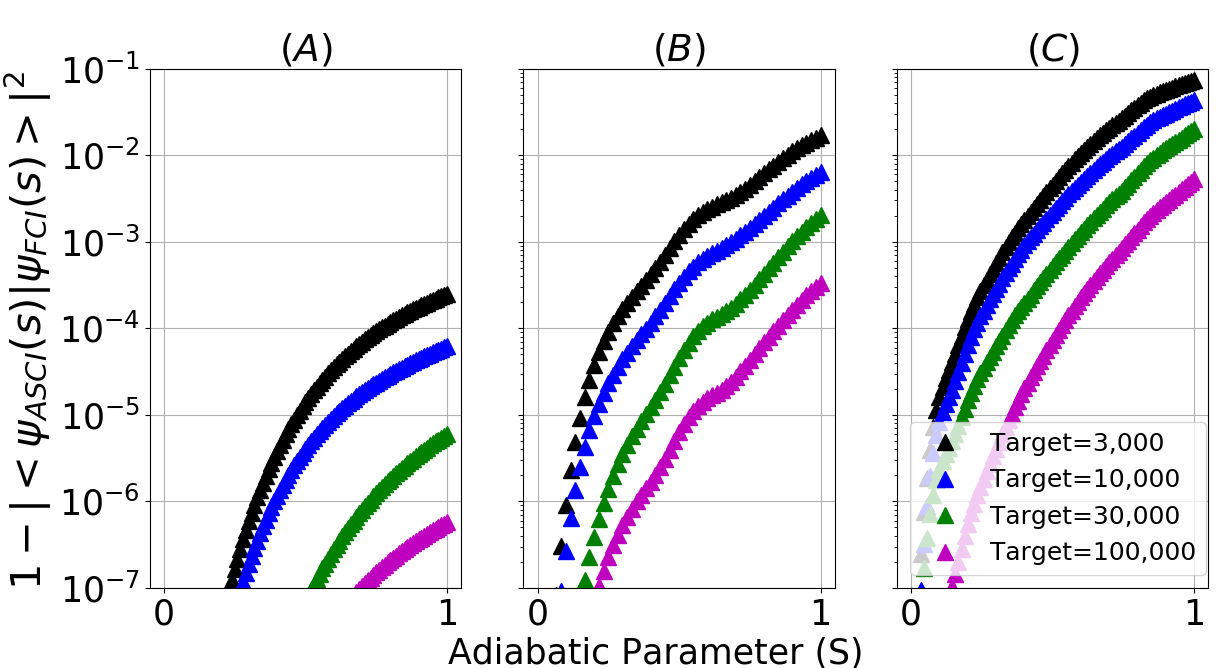}
\caption{ASCI performance in accurately recreating evolved wavefunction for various target sizes (denoted by "target") with a 12 site model at half-filling with a hopping term $t=1$. (A) 2D lattice with periodic boundary conditions, $U=1, \epsilon=0$. (B) 2D periodic lattice $U=3, \epsilon=0$. (C) 1D lattice with open boundary conditions $U=3,\epsilon=1$. The error at each time step represents one minus the inner product between the wavefunction constructed using ASCI states and the wavefunction constructed using FCI. The core size is $2\%$ of target size. }
\label{f:hubbard_asci_vs_fci_overlap_error}
\end{figure}

\begin{figure}[htb]
\centering
\includegraphics[width=\linewidth]{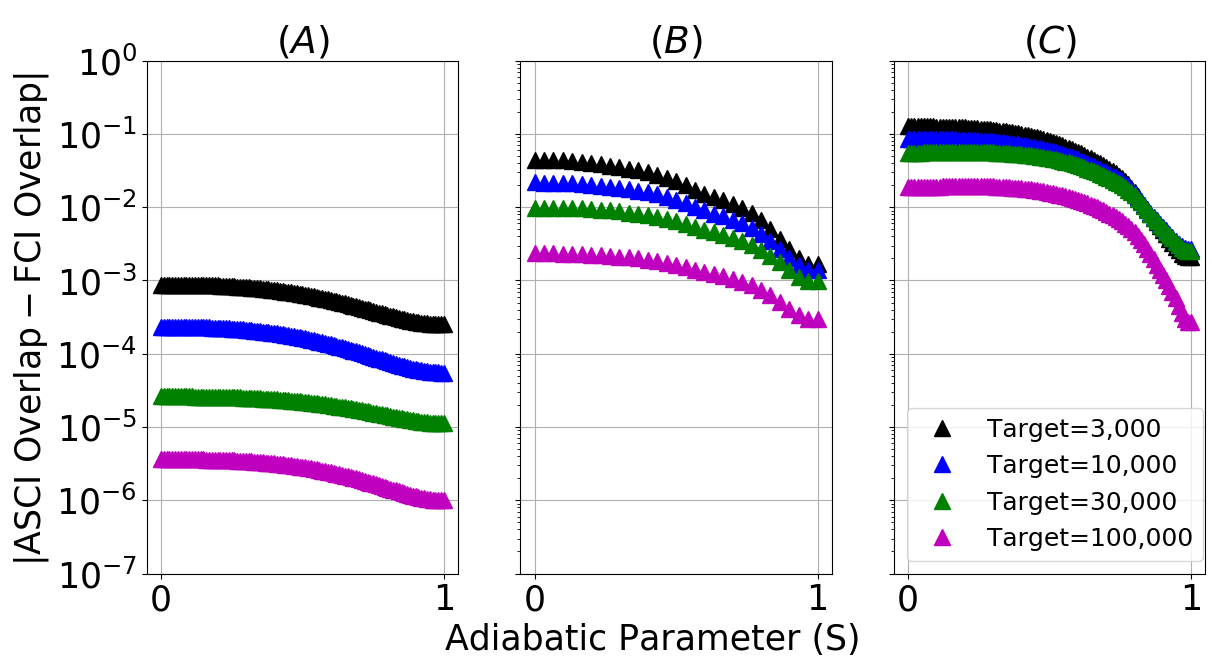}
\caption{ASCI performance for various target sizes (denoted by "target") in the case of a 12 site lattice at half-filling and hopping term $t=1$. (A) 2D lattice with periodic boundary conditions, $U=1, \epsilon=0$. (B) 2D periodic lattice $U=3, \epsilon=0$. (C) 1D aperiodic lattice $U=3,\epsilon=1$. The error at each time step represents the difference between the evolved state's squared overlap with the target state  predicted using ASCI states and the overlap predicted in the full CI case. The core size is $2\%$ of target size. The error decreases over time due to all squared overlaps converging to a common limit of 1.}
\label{f:hubb_asci_ts_error}

\end{figure}

We repeat the error analysis from the molecular set for a set of Hubbard model benchmarks. Firstly, we calculated the squared overlap between the evolved state constructed with ASCI states and the evolved state constructed using full CI at each time step, as seen in \Fig{hubbard_asci_vs_fci_overlap_error}. Secondly, We replicated the calculations for error in predicted overlap from using an ASCI-truncated space. Once more, using $\sim 10^5$ determinants appears to be accurate enough for the purposes of our Hubbard simulations, as can be demonstrated in the three test cases displayed in \Fig{hubb_asci_ts_error}. 

The error is largest specifically for high values of $U$ and low values of $\epsilon$). The worst error in the predicted energy gap (as a percentage) from an ASCI simulation is about $2\%$ for a sufficiently large target size. 

More detailed analysis of overlap and energy gap error for both the molecular and Hubbard model sets may be found in the Appendix.

\subsection{Assessing the adiabatic nature of our protocol}
\begin{figure}
 \centering
\includegraphics[width=\linewidth]{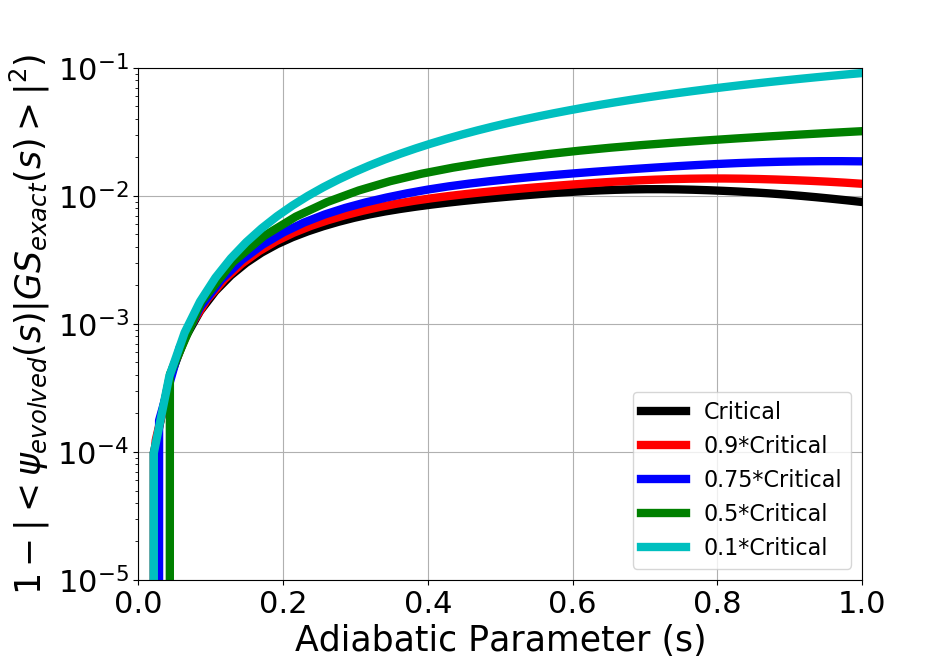}
\caption{A plot of the distance between the evolved state and the instantaneous ground state at each time step in the adiabatic evolution for different preparation times. F\textsubscript{2}, cc-pVTZ basis, target size of $2.5*10^5$, and core size $5*10^3$. }

\label{f:F2_crit_time_errors}
\end{figure}

Another important aspect of assessing our protocol lies in determining just how adiabatic it truly is - that is, how close the prepared state is to the instantaneous ground states throughout the evolution. To assess this, we plotted the squared magnitude of the component of the evolved state that was orthogonal to the true ground state at each time step, as shown in \Fig{F2_crit_time_errors}. Preparing the state at the critical time resulted in the evolved state having at least a $99\%$ squared overlap with the instantaneous ground state not only at the final time step, but throughout the entire evolution. This held true for both the molecular and Hubbard model cases that we examined. 

\subsection{Assessment of critical times of adiabatic evolution: a proxy for quantum resource estimate}
\subsubsection{Molecular Set}
\begin{figure}
\centering
\begin{subfigure}{}
   \includegraphics[width=\linewidth]{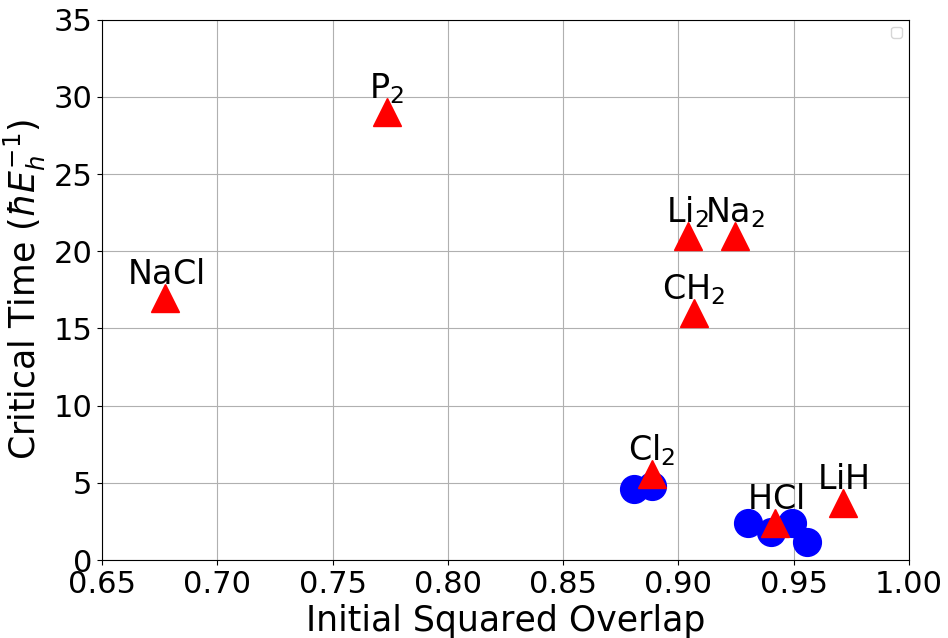}
   \label{fig:asci_overlap_vs_time_dz}
\end{subfigure}
\hfill
\begin{subfigure}{}
   \includegraphics[width=\linewidth]{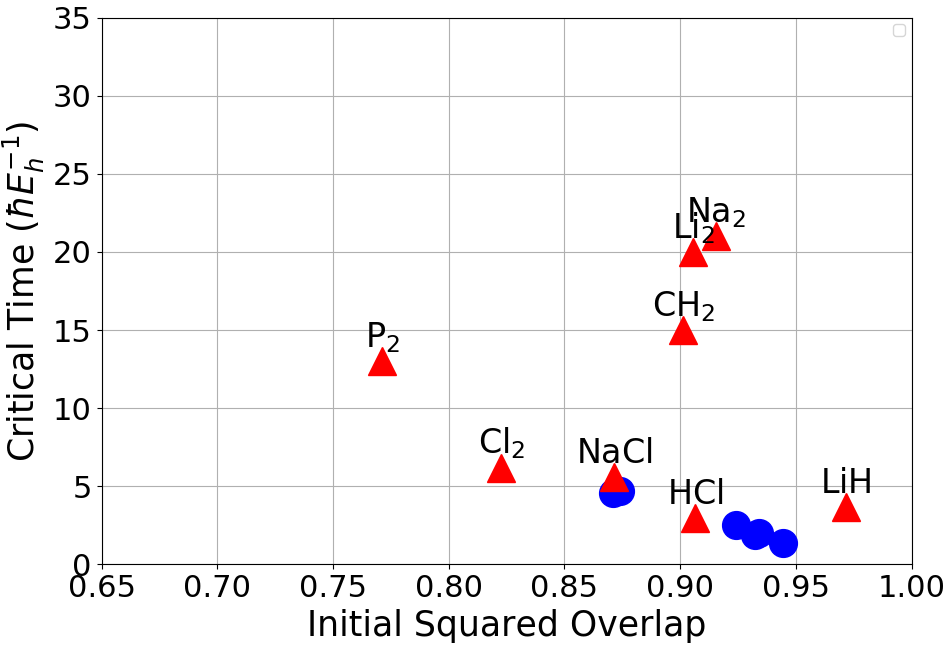}
   \label{fig:asci_overlap_vs_time_tz}
 \end{subfigure}
\caption{Critical time (of adiabatic evolution) required to achieve 99$\%$ squared overlap with target state vs the initial squared overlap with the target state. Red, triangular data points are labeled, blue circular data points are unlabeled. \textbf{TOP}: uses ASCI determinants, the cc-pVDZ basis, and a target size of $10^5$ with an equal core size. \textbf{BOTTOM}: uses ASCI states, the cc-pVTZ basis, and a target size of $10^6$ with a core size that is 2$\%$ of the target size.}
\label{f:mol_overlap_vs_time}
\end{figure}

For the molecular set, we find the initial overlap and the critical time required for state preparation using  a truncated determinant space of ASCI determinants and the cc-pVDZ and cc-pVTZ orbital basis sets. The results are displayed in \Fig{mol_overlap_vs_time} and Table~\ref{table2}. 
For both basis sets, we observe some correlation  between the size of the initial overlap with the target ground state and the time required to prepare the target ground state. With the exception of a few molecules like NaCl and P\textsubscript{2}, the majority of molecular data points do not change significantly between the two orbital basis sets, with very minor changes in initial overlap and almost no changes in critical time. 

\begin{table*}
\begin{center}
\begin{tabular}{ |c|c|c|c|c|c|c|c|c|c|c|c|c|c|c|c|c|c|c|c|c } 
 \hline
 \textbf{Molecule} & H\textsubscript{2}O & LiH & F\textsubscript{2} & HF & HCl & Li\textsubscript{2} & LiF & N\textsubscript{2} & Cl\textsubscript{2} & P\textsubscript{2} & NaCl & Na\textsubscript{2} & CH\textsubscript{4} & CH\textsubscript{2}\\ 
 \hline
 Full CI (cc-pVDZ)\\
 \hline
 \textbf{Spatial Orbitals} & 17 & 19 &  14 &  17 & 17  & 26  & 17  & 14 & 14 & 15 & 17 & 26 & 16 & 20\\ 
 \textbf{Initial Squared Overlap} & 0.952 & 0.973 & 0.915 & 0.958 & 0.942 & 0.904 & 0.983 &  0.931 & 0.951 & 0.867 & 0.918 & 0.925 & 0.966 & 0.918\\ 
 \textbf{Minimum Gap} &  0.286 & 0.115 & 0.140 & 0.379 & 0.281 & 0.049 & 0.238 & 0.309 & 0.130 & 0.114 & 0.166 & 0.038 & 0.410 & 0.068\\
 \textbf{Critical Time (overlap)} & 1.8 & 3.7 & 4.7 & 1.2 & 2.3 & 23 & 2.0 & 4.5 & 3.8 & 14 & 19 & 21 & 2.3 & 15\\
\hline
 ASCI (cc-pVDZ)\\
 \hline
 \textbf{Spatial Orbitals} & 23 & 18 &  26 &  18 & 18  & 26  & 26 & 26 & 26 & 26 & 26 & 26 & 33 & 23\\ 
 \textbf{Initial Squared Overlap} & 0.940 & 0.972 & 0.888 & 0.956 & 0.942 & 0.904 & 0.949 & 0.881 &  0.889 & 0.774 & 0.677 & 0.925 & 0.930 & 0.907\\ 
 \textbf{Minimum Gap} & 0.277 & 0.116 & 0.125 & 0.377 & 0.266 & 0.048 & 0.224 & 0.291 & 0.102 & 0.050 & 0.397 & 0.038 & 0.424 & 0.066\\
 \textbf{Critical Time (overlap)} & 1.8 & 3.7 & 4.8 & 1.2 & 2.4 & 21 & 2.4  & 4.6 & 5.6 & 29 & 17 & 21 & 2.4 &  16\\
 \textbf{Critical Time (energy)} & 6.0 & 6.2 & 12.2 & 4.8 & 7.9 & 21.8 & 7.9 & 11.9 & 9.1 & 31.9 & 22.1  &  21.7& 7.4 & 19\\
 
 \hline
 ASCI (cc-pVTZ)\\
 \hline
 \textbf{Spatial Orbitals} & 57 & 43 & 58 & 43 & 43 & 58 & 58& 58 & 58 & 58 & 58 & 58 & 64 & 57 \\
 \textbf{Initial Squared Overlap} & 0.932 & 0.972 & 0.874 & 0.944 & 0.907 & 0.906 & 0.934 & 0.871 & 0.823 & 0.771 & 0.872 & 0.916 & 0.924 & 0.901\\
 \textbf{Minimum Gap} & 0.279 & 0.116 & 0.129 & 0.391 & 0.275 & 0.048 & 0.241 & 0.316 & 0.148 & 0.106 & 0.173 & 0.039 & 0.408 & 0.017\\
 \textbf{Critical Time (overlap)} & 1.9 & 3.7 & 4.7 & 1.4 & 2.9 & 20 & 2.0 & 4.6 & 6.2 & 13 & 5.6 & 21 & 2.5 & 15\\
 \hline
\end{tabular}
\end{center}
\caption{\label{table2} A display of the simulation results for the part of the molecular set that was tested with truncated and untruncated determinant spaces and different orbital basis sets. Energies displayed in Hartree ($E_h$), times displayed in atomic units ($\hbar E_h^{-1}$). The initial squared overlap refers to the square norm of the inner product between the initial ground state and the target ground state. The minimum gap refers to the minimum difference between the ground and first excited state energies over the course of the evolution. Two critical times are included, one specifying the minimum amount of time required to evolve the state so that its squared overlap with the target state is at least 0.99, and one specifying the amount of time required to evolve the state so that its expectation energy is within chemical accuracy of the ground state energy.}
\end{table*}

\begin{table*}
\begin{center}
\begin{tabular}{ |c|c|c|c|c|c|c|c|c|c|c|c|c|c|c|c|c|c|c|c|c } 
 \hline
 \textbf{Molecule} & CO\textsubscript{2} & ClF & SiH\textsubscript{4} & NH\textsubscript{3} & SiH\textsubscript{2} & H\textsubscript{2} & LiNa & NaH\\ 
 \hline
 ASCI (cc-pVTZ)\\
 \hline
 \textbf{Spatial Orbitals} & 64 & 58 & 64 & 64 & 57 & 28 & 58 & 43\\
 \textbf{Initial Squared Overlap} & 0.855 & 0.871 & 0.911 & 0.926 & 0.891 & 0.982 & 0.911 & 0.964\\
 \textbf{Minimum Gap} & 0.329 & 0.134 & 0.335 & 0.266 & 0.049 & 0.393 & 0.043 & 0.101\\
 \textbf{Critical Time (overlap)} & 5.1 & 4.1 & 3.9 & 2.2 & 12 & 1.4 & 20 & 4.7\\
 \hline
\end{tabular}
\end{center}
\caption{\label{table3} A display of the simulation results for the part of the molecular set for which we only used the cc-PVTZ orbital basis. Energies displayed in Hartree ($E_h$), times displayed in atomic units ($\hbar E_h^{-1}$). The initial squared overlap refers to the square norm of the inner product between the initial ground state and the target ground state. The minimum gap refers to the minimum difference between the ground and first excited state energies over the course of the evolution. The critical time  specifies the minimum amount of time required to evolve the state so that its squared overlap with the target state is at least 0.99.}
\end{table*}

\begin{figure}[htb]
\centering
\includegraphics[width=\linewidth]{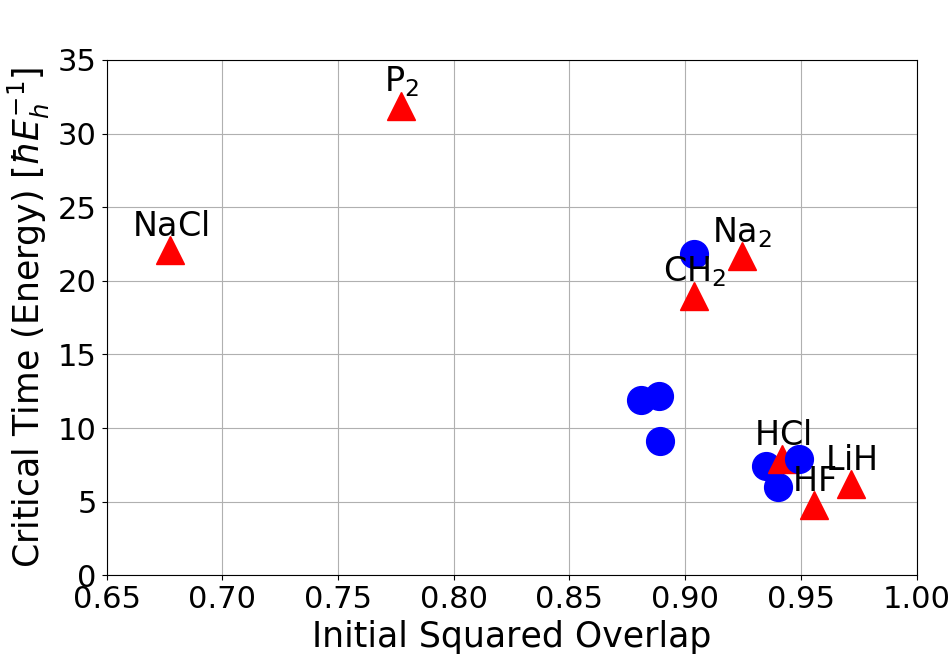}
\caption{Initial overlap $|\bra{\psi(s=0)}\ket{\psi_{target}}|^2 $ with the target ground state and the minimal evolution runtime (atomic units) required for the prepared state's expected energy to be within 1.6 mHa of the true ground state energy (chemical accuracy). Using ASCI-generated states and cc-pVDZ basis, core size = 2\% target size. Red markers are used for labeled data points, blue for unlabeled.}
\label{f:energyerrorvtime_ccpvdz}
\end{figure}

\begin{figure}[htb]
\centering
\includegraphics[width=\linewidth]{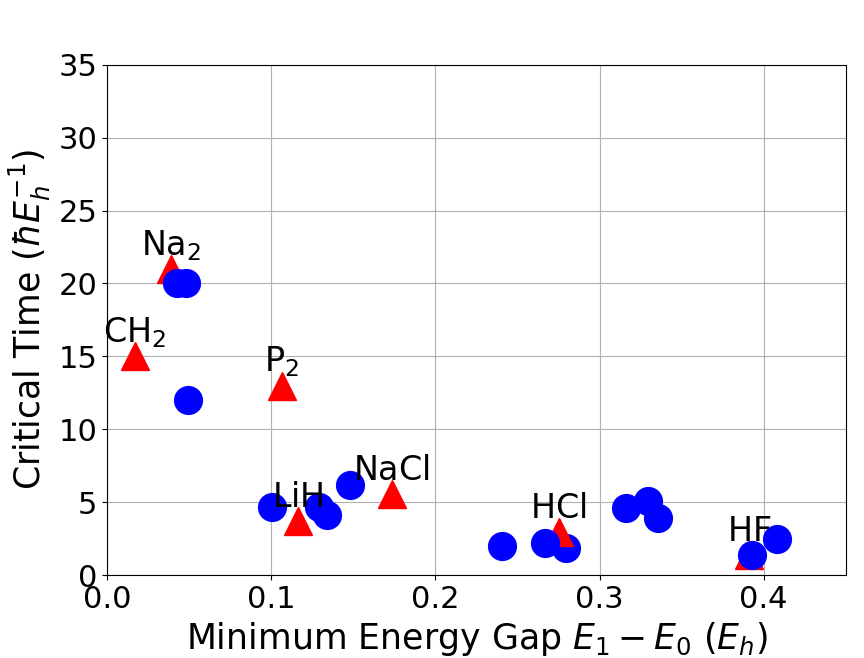}
\caption{Minimum gap between the first excited and ground state energies ($E_h$) plotted against the minimal evolution runtime (atomic units) required for the prepared state to have an overlap of at least 0.99 squared overlap with the target state. Using ASCI-generated states and cc-pVTZ basis, core size = 2\% target size. Red markers are used for labeled data points, blue for unlabeled.}
\label{f:mingap_vs_time_triple_zeta}
\end{figure}

We also test the relationship between the initial overlap with the target ground state and the critical time required to prepare a state whose expectation energy is within chemical accuracy of the true ground state energy. This was done with ASCI using the cc-pVDZ orbital basis. The time required for state preparation with this new criterion was consistently higher than the time required to achieve 99$\%$ overlap, but the overall distribution of the data points, though scaled, was not substantially changed as can be seen by comparing \Fig{energyerrorvtime_ccpvdz} and the upper panel of \Fig{mol_overlap_vs_time}.
We  plot the critical time required to prepare the ground state against the minimum energy gap in \Fig{mingap_vs_time_triple_zeta} with data points from all the molecules simulated using the cc-pVTZ basis. We would expect the shape of the data distribution to resemble an inverse-square curve based on the denominator of Eq. \eqref{eq:critical_time_constraints} if we ignore effects from the numerator.

\subsubsection{Hubbard Set}
\begin{figure}[htb]
\centering
\includegraphics[width=\linewidth]{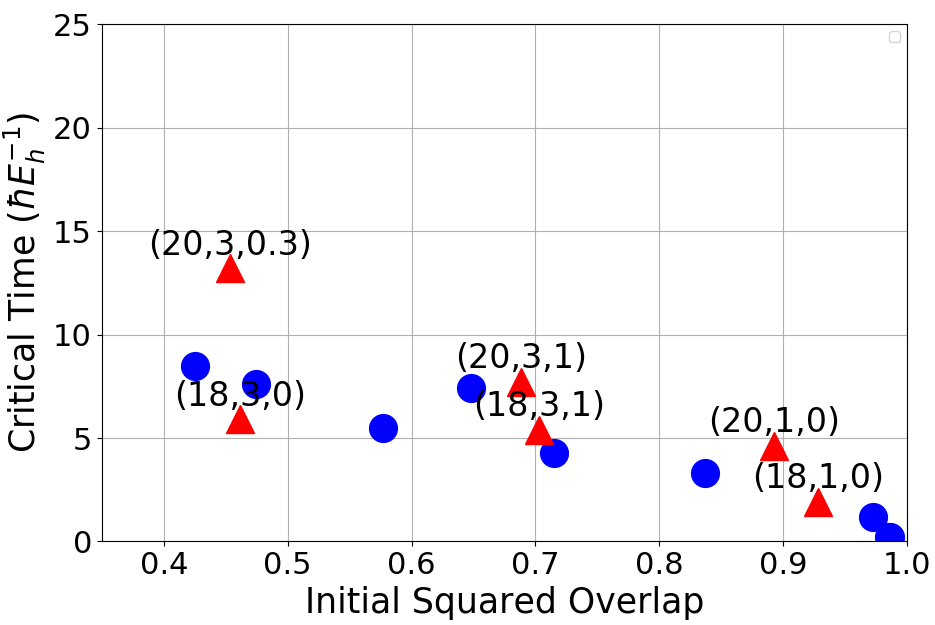}
\caption{Initial overlap $|\bra{\psi(s=0)}\ket{\psi_{target}}|^2 $ with the target ground state versus the minimal evolution runtime (atomic units) required for the prepared state to have at least a $99\%$ overlap with the target ground state. Using ASCI-generated states and  core size = 2\% target size. Red markers are used for labeled data points, blue for unlabeled, with label formatting being (number of sites, U, $\epsilon$).}
\label{f:hubb_overlap_vs_time}
\end{figure}

We also examine the relationship between the initial overlap and the state preparation time for a set of Hubbard model parameters, the results for which are plotted in \Fig{hubb_overlap_vs_time}. Once again we find the expected effect that increasing the initial overlap with the target ground state decreases the amount of time required to prepare it. The correlation in this case appears to be much more clear than the molecular case, with all of the different Hubbard model data points falling along the same downward-sloping, line-like cluster. For both site counts, increasing U (holding t fixed at 1) decreases the initial overlap and increases the required preparation time. Increasing $\epsilon$ initially increases then gradually decreases the critical time, while increasing the initial overlap with an indeterminate effect on the minimum gap size.

\begin{table*}
\begin{center}
\begin{tabular}{ |c|c|c|c|c|c|c|c|c|c|c|c|c|c|c| } 
 \hline
 \textbf{(sites,U,$\epsilon$)} & (18,1,0) & (18,1.5,0) & (18,2,0) & (18,2.5,0) & (18,3,0) & (18,3,0.3) & (18,3,1) & (18,3,3)\\ 
 \textbf{Initial Squared Overlap} & 0.928 & 0.837 & 0.715 & 0.577 & 0.461 & 0.474 & 0.703 & 0.987 \\
 \textbf{Minimum Gap} & 0.584 & 0.549 & 0.571 & 0.519 & 0.461 & 0.526 & 0.660 & 3.459 \\
 \textbf{Critical Time (overlap)} & 1.9 & 3.3 & 4.3 & 5.5 & 5.9 & 7.6 & 5.4 & 0.2\\
 \hline

 \hline
 \textbf{(sites,U,$\epsilon$)} & (20,0.5,0) & (20,1,0) & (20,2,0) & (20,3,0) & (20,3,0.3) & (20,3,1) & (20,3,3) & --\\ 
 \textbf{Initial Squared Overlap} & 0.973 & 0.892 & 0.648 & 0.425 & 0.453 & 0.688 & 0.986 & -- \\
 \textbf{Minimum Gap} & 0.268 & 0.248 & 0.244 & 0.501 & 0.253 & 0.480 & 3.399 & --\\
 \textbf{Critical Time (overlap)} & 1.2 & 4.6 & 7.4 & 8.5 & 13.2 & 7.7 & 0.22  & --\\
 \hline
\end{tabular}
\end{center}
\caption{A display of the results for the Hubbard model set using ASCI truncated determinant space. Presented are the squared overlap between the starting and target ground states, the minimum time ($\hbar E_h^{-1}$) required to prepare a state with at least 99$\%$ squared overlap with the target ground state, and the minimum gap ($E_h$) between the first excited and ground states over the course of the evolution. \label{table4}}
\end{table*}

\subsection{Number of valence electrons and sensitivity of preparation time to initial overlap in the molecular set}
\label{s:valence_trends}
\begin{figure}[htb]

\centering
\includegraphics[width=\linewidth]{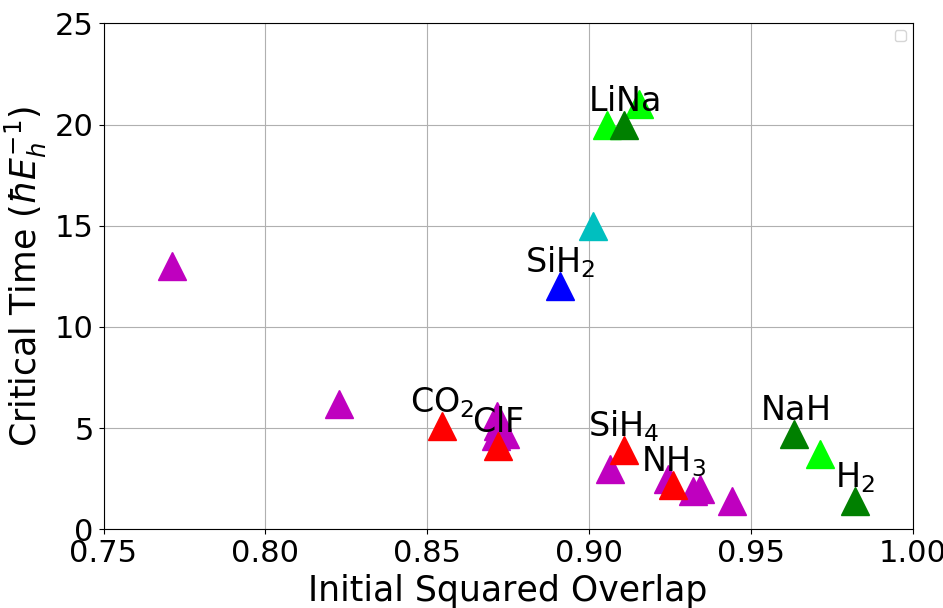}
\caption{A plot of the initial overlap $|\bra{\psi(s=0)}\ket{\psi_{target}}|^2 $ with the target ground state and the minimal evolution runtime (atomic units) required so that the final squared overlap with the target state is at least 0.99. Using ASCI-generated states and cc-pVTZ basis. Purple/red markers are used to denote unlabeled/labeled molecules with at least 8 valence electrons, cyan/dark blue markers used for unlabeled/labeled molecules with six valence electrons, and light/dark green markers used for unlabeled/label molecules with 2 valence electrons. All labeled molecules are the newly added ones. }

\label{f:adding_tz_molecules}
\end{figure}

To better understand some of the observed trends, we used a larger set of molecules for the cc-pVTZ basis (adding CO\textsubscript{2}, SiH\textsubscript{4}, NH\textsubscript{3}, ClF, SiH\textsubscript{2}, LiNa, NaH, and H\textsubscript{2}). One such trend we noticed was that the data points appeared to be forming separate linear clusters based on the molecule's number of valence electrons. As can be seen in the final plot of \Fig{adding_tz_molecules}, molecules with only two valence electrons seem to fall along the steepest cluster, all molecules with eight or more valence electrons fall along the least steep cluster on the bottom, and those with  six valence electrons fall somewhere in between.

Additionally, when we plotted initial overlap against time for multi-determinant initial ground states (see Section \ref{s:multi_determinants}), these clusters were preserved, with the molecular data points simply "sliding down" the clusters as the initial overlap increased and state preparation time decreased. 

\subsection{Minimum gap location over the course of the evolution}
\label{s:min_gap_location}
\subsubsection{Molecular Set}
\begin{figure}
 \centering
\includegraphics[width=\linewidth]{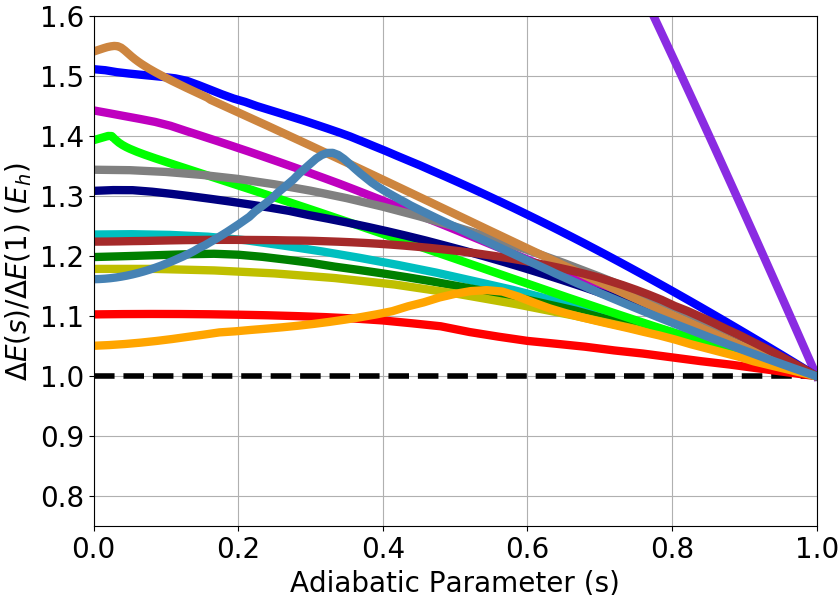}
\caption{A plot of the ratio of the energy gap ($E_h$) at a particular time step to the energy gap at the final time step over the course of the evolution for all initial fourteen molecules. Light purple curve belongs CH\textsubscript{2}, convex curve peaking at 3.0 at the beginning. Using ASCI-generated determinants and cc-pVTZ basis. }

\label{f:mol_min_gap_location}
\end{figure}

Plotting the size of the gap between the ground and first excited energies over the course of the adiabatic evolution  in \Fig{mol_min_gap_location} shows that the gap reaches its minimum at the final time step for all our tested molecules. Investigating this for the eight additional molecules tested with the cc-pVTZ basis, we found precisely the same trends, with the ratio lying between 1 and 1.6 just like for almost all molecules in \Fig{mol_min_gap_location}. Knowing the location of the minimum gap could prove especially helpful (as is the fact that the gap is significantly larger than its minimum for most of the evolution) since this could allow us to speed up the state preparation process by evolving our Hamiltonian faster at the beginning of the evolution and slower towards the end. We investigate this approach in Section \ref{s:nonlinear_interp}. 

\subsubsection{Hubbard Set}
\begin{figure}
 \centering
\includegraphics[width=\linewidth]{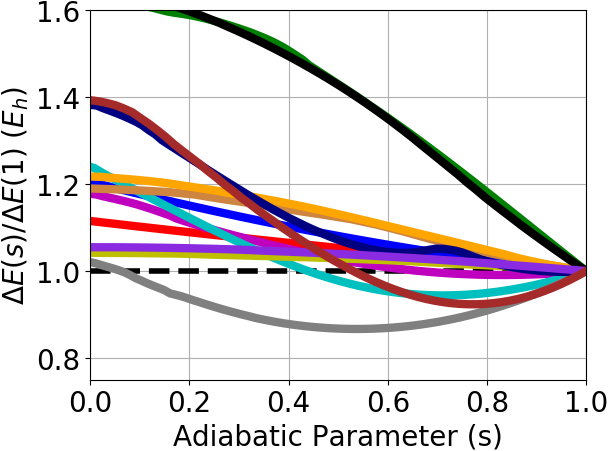}
\caption{A plot of the ratio of the energy gap ($E_h$) at a particular time step to the energy gap at the final time step over the course of the evolution for the Hubbard models examined. ASCI-truncated determinant set used.}
\label{f:hub_min_gap_location}
\end{figure}

This trend of the minimum gap occurring in the final time step does not hold for all of the Hubbard models that we examined. As can be seen in \Fig{hub_min_gap_location}, in some cases the minimum gap occurred as early as the middle of the evolution. Furthermore, the shape of the gap in those cases takes on a somewhat unusual, concave form. The Hubbard models in those cases were all models where $U$ was high and the Hartree-Fock state's overlap with the target ground state was very low.

\subsection{Speeding up adiabatic state preparation through multi-determinant starting ground states.}
\label{s:multi_determinants}
\subsubsection{Molecular Set}
\begin{figure}
 \centering
\includegraphics[width=\linewidth]{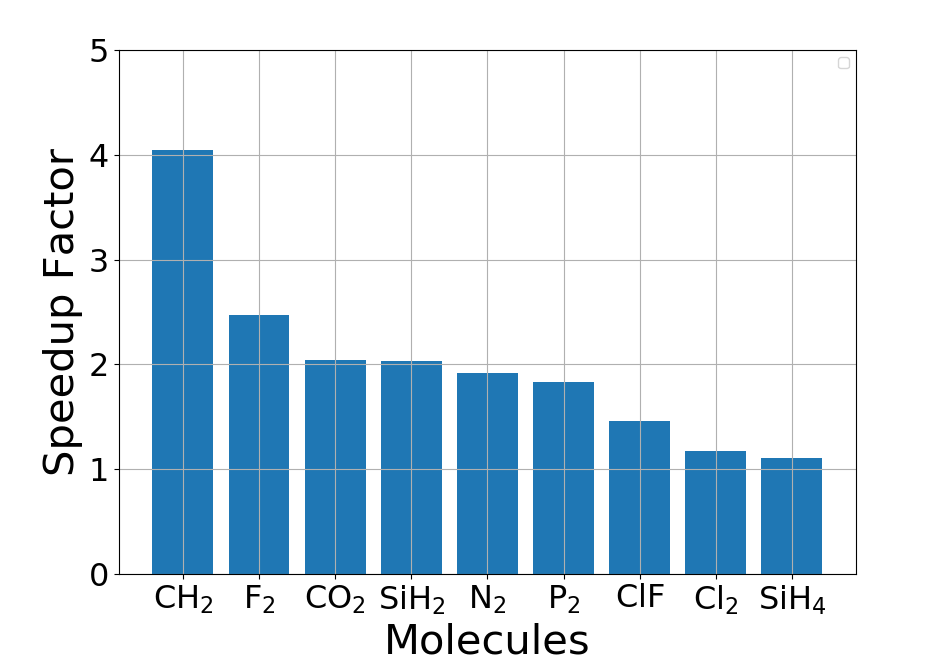}
\caption{Factor by which the critical preparation time is reduced for each molecule by allowing CASCI-like initial ground state. Only molecules listed for which there was a speedup of at least 1.1. The critical time for each molecule was calculated using an ASCI-truncated space with $10^6$ target size, $2*10^4$ core size, and the cc-pVTZ basis.}
\label{f:multi1_mol_speedups}
\end{figure}

\begin{figure}
 \centering
\includegraphics[width=\linewidth]{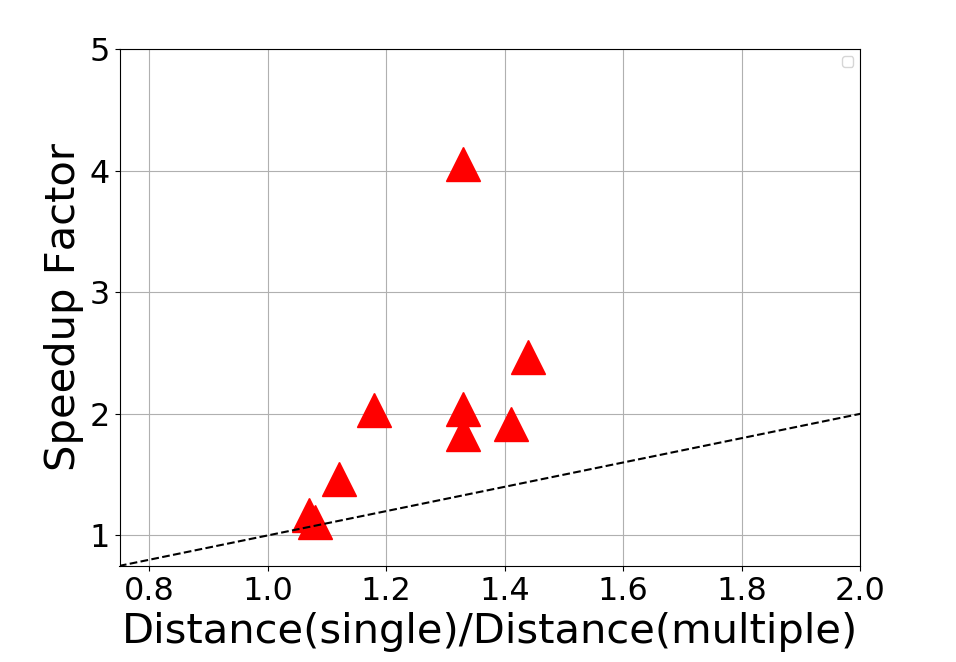}
\caption{Factor by which the critical preparation time is reduced vs factor by which distance to target state is reduced by having the initial ground state be a multi-determinant, CASCI wavefunction. Distance is defined as $1-|\bra{\psi_{\text{initial}}}\ket{\psi_{\text{target}}}|^2 $. The critical time for each molecule was calculated using an ASCI-truncated space with $10^6$ target size, $2*10^4$ core size, and the cc-pVTZ basis. The data points here correspond to the molecules from \Fig{multi1_mol_speedups}.}
\label{f:multi1_mol_speedups_vs_distance}
\end{figure}

We investigated the impact of having an initial Hamiltonian whose ground state was made up of several Slater determinants rather than simply one (the Hartree-Fock state). To this end, we use CASCI wavefunctions with an increased number of orbitals to complete the lowest valence shell.The corresponding initial Hamiltonian in this case keeps all terms connecting determinants in the CASCI active space, while setting all other off-diagonal elements to zero. As can be seen in \Fig{multi1_mol_speedups}, this reduces the required preparation time by a factor of 2-4 for some the molecules. Plotting the factor by which the preparation time was reduced against the factor by which "distance" to the target state (measured as $1-|\bra{\psi_{initial}}\ket{\psi_{target}}|^2$) was reduced shows that the molecules whose preparation time did not meaningfully decrease also did not have a meaningful change in initial overlap as shown in \Fig{multi1_mol_speedups_vs_distance}).

\begin{figure}
 \centering
\includegraphics[width=\linewidth]{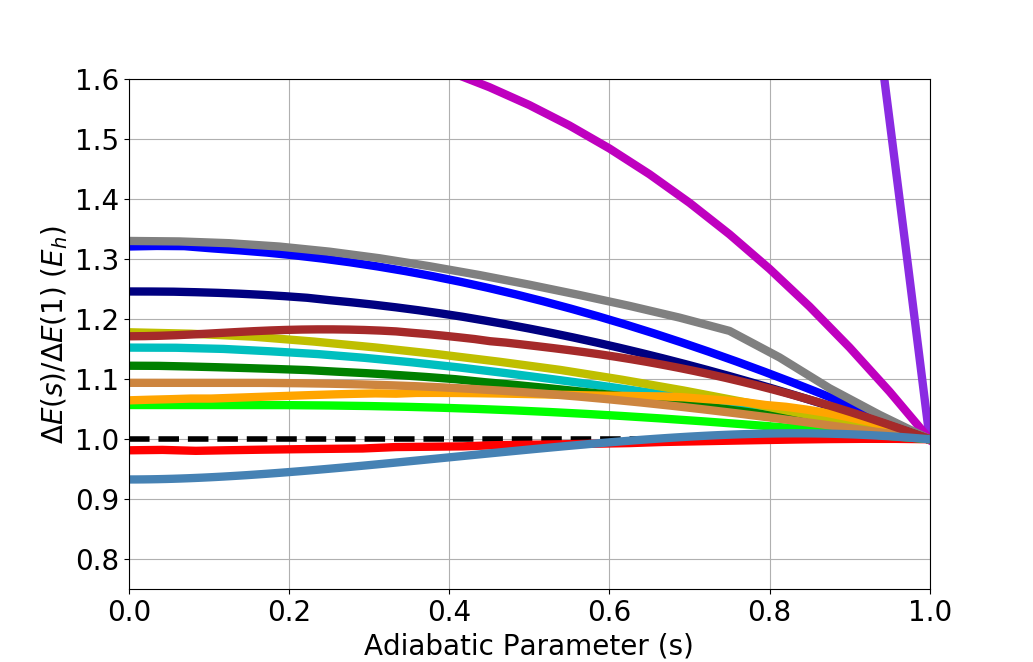}
\caption{A plot of the ratio of the energy gap ($E_h$) at a particular time step to the energy gap at the final time step over the course of the evolution for all tested molecules. Here, the initial ground state is made up of multiple determinants, as described in Section \ref{s:multi_determinants}.}

\label{f:multi_mol_min_gap_location}
\end{figure}

The trends observed in Section \ref{s:valence_trends} continue here, with the molecular data points forming the same distinct clusters based on the number of valence electrons the molecule has, and the data points falling into the region of the cluster corresponding to their initial overlap. 
Additionally, it appears that the minimum energy gap no longer takes place at the end of the evolution for some molecules once a multi-determinant initial ground state is used. Nevertheless, it still appears to take place at the end for the super-majority of them, and the cases where it does not are the cases where the size of the gap barely changes at all as the system is evolved. Additionally, as shown in \Fig{multi_mol_min_gap_location}, the gap undergoes a significantly smaller change compared to what's seen in \Fig{mol_min_gap_location}.

\subsubsection{Hubbard Set}
In order to make a similar attempt at speeding up the preparation of our Hubbard model ground states, we found the most frequently occupied orbitals (besides the occupied orbitals in the Hartree-Fock state) for the ground state of each model and used multi-determinant initial ground states where those were the occupied orbitals. We then examined how using these initial ground states affected the critical time of the models whose preparation time exceeded 5 $\hbar^{-1}E_h$. This did not produce any significant speedups, with the most noteworthy speedup factors being around 1.5 for models with open boundary conditions. Models with periodic boundary conditions not only failed to exhibit meaningful speedups, but did so despite substantial reductions in the distance to the target state ($1-|\bra{\psi_{\text{initial}}}\ket{\psi_{\text{target}}}|^2)$ by factors of 1.2-1.5. 
\subsection{Speeding up adiabatic state preparation through non-linear interpolation between Hamiltonians.}
\label{s:nonlinear_interp}
Given that we know the minimum gap occurs at the end of all the molecular cases with the Hartree-Fock initial state and most of the Hubbard ones, one potential approach to speeding up the adiabatic preparation process would be to evolve the Hamiltonian more quickly at the beginning of the evolution and slow down as we approach the end. We investigated this approach using a nonlinear, polynomial interpolation between the initial and target Hamiltonians:
$$H(s) = (1-s^c)H_{\text{initial}} + s^c H_{\text{target}},\ 0 < c < 1$$
While making the interpolation non-linear did provide an advantage, said advantage was not significant in any of the molecules we investigated, lowering the critical time of preparation by $\sim 0.1 $ a.u. at the absolute most. Furthermore, the optimal value of $c$ for these slight improvements was $c \approx 0.9$ across all the tested cases, meaning the optimal interpolation was nearly linear. In all likelihood, the lack of significant gains stemmed from the energy gap not undergoing particularly large changes over the course of the adiabatic evolution in the physical systems that we examined. It is possible that non-linear interpolation yields useful speedups for energy gaps that undergo more drastic changes during the evolution.

\section{Discussion}
\subsection{ASCI accuracy}
One of the key advances provided in this paper is a powerful method for treating the adiabatic evolution of large systems. Using the ASCI method of selecting important determinants for constructing the ground and first excited states reduces the required simulation resources without a substantial trade-off in accuracy.

The initial, relatively high proximity to the target ground state enables us to only have to find the states important for the construction of the final ground and excited states in order to find the determinants important for the time dynamics, without having to repeat the ASCI protocol for the Hamiltonian at earlier time steps. Additionally, the prepared wavefunction stays overwhelmingly in the instantaneous ground state at every step in the evolution and so is not significantly modified by us keeping only the important determinants for the first two eigenstates.

For both the molecular and Hubbard model sets, $10^6$ determinants appear to be enough to predict the initial and final squared overlaps with the target ground state within at most $10^{-2}$ and $10^{-3}$ relative error, respectively. At most another $10^6$ determinants for the excited state suffice to determine the energy gap within a few percent. The addition of these extra determinants to find the excited energy does not have a noticeable impact on the accuracy of the predicted overlap between the evolved state and the target ground state. This is likely because the initial ground state was close enough to the target ground state such that even substantial errors in the minimum gap size would not meaningfully affect the evolution in the state.  

Also, it appears that the critical preparation time of molecular ground states may be usually determined fairly accurately with only a small number of spatial orbitals (using just the cc-pVDZ basis for example) and even provides a close estimate of the initial overlap. For reference, consider \Fig{mol_overlap_vs_time}, where for almost all the molecules (with the notable exception of NaCl and P\textsubscript{2}) going from the cc-pVDZ to the cc-pVTZ basis (top to bottom plots) only slightly moves the molecular data points to the left and barely changes their height. That is, it only slightly changed the predicted initial overlap and had almost no effect on the predicted critical preparation time.
\subsection{Methods for speeding up adiabatic state preparation}
For molecules, it appears that having the initial ground state be a CASCI wavefunction produces substantial reductions in the required state preparation time, generally by a factor of 2-4, for wavefunctions produced by allowing enough active orbitals to close the first valence shell. Speedups may be even more substantial by having starting CASCI wavefunctions from starting Hamiltonians with even more active orbitals. For both the molecules and the Hubbard models, it appeared that allowing multi-determinant initial ground states introduced a speedup factor proportional to the factor by which the distance in state space to the target state was reduced. Generally, only a small number of orbitals needed to be added in order for these speedups to be substantial. As such, this might be a useful means of quickening state preparation in other, more complicated physical systems.

Running simulations investigating non-linear interpolation between starting and target Hamiltonians resulted in little-to-no improvement in the state preparation time. However, this was very likely a consequence of the fact that the energy gaps in all of the systems that we examined underwent very little proportional change over the course of the evolution. It may be useful to investigate the benefits of such interpolations when the gap undergoes much more drastic changes, such as with the non-equilibrium geometries of methylene discussed in \cite{veis2014} where non-linear interpolation yielded significant speedups.

Additionally, knowledge that the minimum energy gap appears at or near the end of the evolution when starting from the Hartree-Fock state for a large array of physical systems (and all the molecular systems we examined) might prove useful for future efforts to reduce the state preparation time. This might help guide the choice of optimal interpolation method between Hamiltonians in cases where this offers some speedup. Additionally, this information might be helpful in the context of annealing as seen in  \cite{marshall2019, gonzales2020}, if annealing hardware can implement the gates necessary to prepare and evolve the ground state of such physical systems. 

\subsection{Useful Trends}
It is important to note that none of the parameter changes that we experimented with across simulations (whether it was the kind of physical systems, the number of valence electrons in molecular systems, the kind of interpolation used or the choice of starting ground state) resulted in an exponential increase in the state preparation time. This suggests that adiabatic state preparation is an effective method for simulation of these physical systems.

Another intriguing trend that we observed in the molecular data seemed to indicate that the number of valence electrons in the molecules affected the dependence of the critical state preparation time on the initial overlap with the target ground state. In particular, increasing the amount of molecular valence electrons up to eight reduced the sensitivity of preparation time to initial overlap. It would be interesting to investigate the cause of this relationship, and see whether it might be useful in either predicting preparation time in advance or have use in speeding up the preparation process for other physical systems.

Within each cluster, the molecular data points seemed to exhibit the expected inverse relationship between the size of the initial squared overlap with the target ground state and the amount of time needed to prepare the target state. This relationship seemed to be roughly linear, though this has no theoretical backing and is only something the figures for the systems we examined seemed to display. The same broad trend was observed for the Hubbard data points as well. 

 While these results have only been demonstrated for a relatively small number of physical systems, the numerical validation of such trends can be taken as at least an indication that quantum computing will prove useful in these kinds of chemistry simulations. 

\section{Conclusion}In this paper we explored the possibility of using adiabatic state preparation over a wide range of systems.  We observed trends for the critical times that are quite promising and do not show an immediate exponential wall that would be prohibitive to run such state preparation algorithms eventually on quantum hardware.

We developed a time dependent algorithm to simulate adiabatic time evolution classically and  have verified, for an array of different physical systems,  that performing time-dynamics simulations of an adiabatic process using an ASCI-truncated space of determinants does not introduce any substantial errors over our test systems.  As a result, this approach can be used to  efficiently simulate much larger systems or those requiring a much larger amount of orbitals to accurately describe than would be possible using FCI.

Additionally, we confirmed the expected negative correlation between initial overlap with the ground state and the time required to prepare said ground state through ASP for several physical systems, as well as confirming this relationship is not exponential for the conditions that we explored.

We found that, during the adiabatic preparation of the molecular ground states we considered, the minimal energy gap always occurred at the end of the evolution if the starting ground state was the Hartree-Fock state. The same was not always true for the Hubbard models that we examined, though in all cases the variation of the energy gap over the course of the evolution was not dramatic (shrinking by only a factor of 3 in the most extreme case). 

We found evidence that substantial speedups in adiabatic preparation can be made by choosing a CASCI wavefunction as the initial ground state instead of just the Hartree-Fock state, illustrating a general scheme for increasing the efficiency of ASP procedures on quantum hardware. Also considered was the use of non-linear interpolation, as suggested previously, but, in the cases treated here, essentially no benefit was found. This was likely due to these particular systems exhibiting a relatively constant change in gap size which, however, is not the case for all potential systems one may be interested in simulating. It would be interesting to see an investigation of these speedup methods for systems with more strongly correlated electrons.
\section*{Acknowledgments:}
C.M.Z. was supported by Basic Energy Sciences, Chemical Sciences, Geosciences and Biosciences Division, as part of the Computational Chemical Sciences Program, of the U.S. Department of Energy under Contract No. DE-AC02-05CH11231. N.M.T., V.K., and S.C. are grateful for support from NASA Ames Research Center and support from the AFRL Information Directorate under Grant No.~F4HBKC4162G001 and from DARPA under IAA 8839 Annex 125. V.K. is thankful for support from NASA Academic Mission Services, Contract No. NNA16BD14C. Calculations were performed as part of the XSEDE computational Project No. TG-MCA93S030. 

\bibliography{refs}

\appendix
\label{s:appendix1}
\section{Error Analysis}

\subsection{Error From Finite Time Steps}
Below is a chart showing the squared overlap of the prepared state with the target ground state over the course of the adiabatic evolution, for the full CI cases of diatomic Phosphorus and Sodium with the specified number of orbitals:
\begin{figure}[H]
\centering
\includegraphics[width=\linewidth]{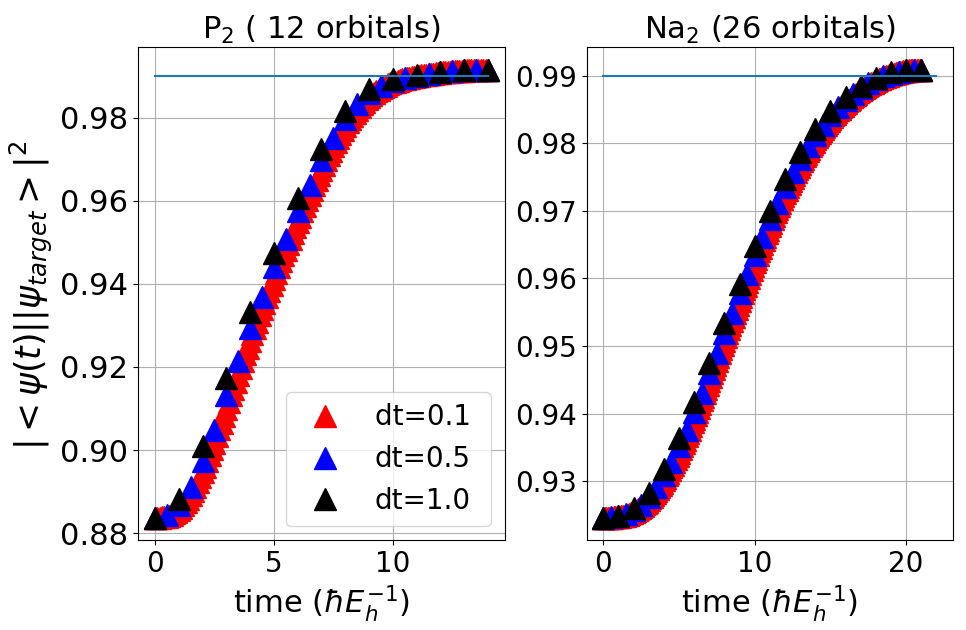}
\caption{Squared overlap of the evolved state  with the target ground state at each time step of a single evolution for three different time step sizes.  }
\label{f:time_step}
\end{figure}

As can be seen in \Fig{time_step}, the squared overlap curve generated using time steps of 0.5 very closely follows the curve generated with time steps of 0.1. By contrast, significant deviations start to occur when we set the time step size equal to 1.
However, for all time step sizes the curves converged to almost exactly the same squared overlap by the end. In fact, the difference between the final overlap using a time step size of 0.5 and using a size of 0.1 was at most only $~ 10^{-5}$ in all the examined molecules. Since the error from finite time steps is a quickly converging one, this indicates that the error in final overlap for $\Delta t =0.1$ is even smaller. 

\subsection{Error From Incomplete Use Of Basis }

\begin{figure}[H]

\centering
\includegraphics[width=\linewidth]{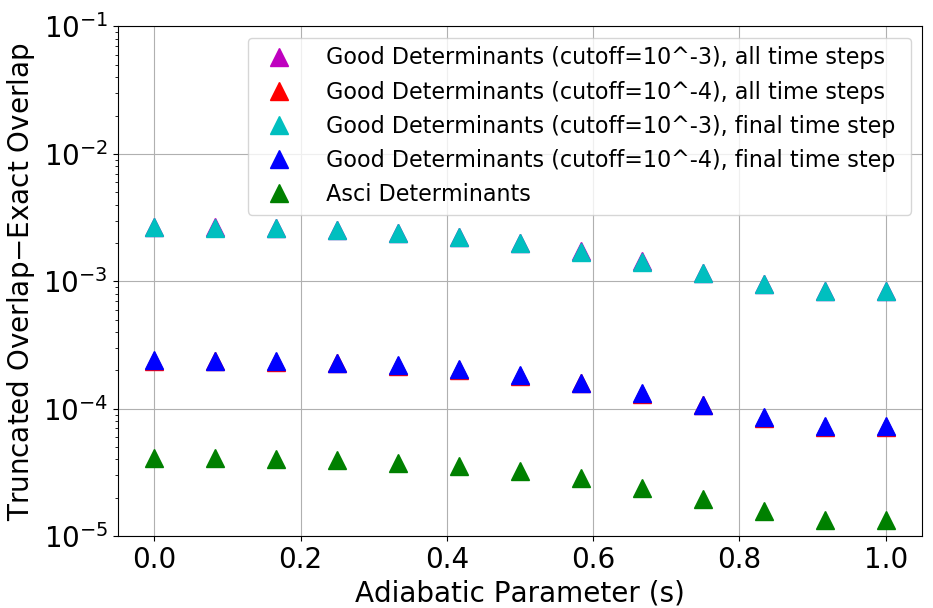}
\caption{Truncation Comparisons for HF (13 spatial orbitals, cc-pVDZ): difference between predicted squared overlap with the target ground state using a truncated state space and using the full state space. Finding good determinants from only the final time step's ground state had a negligible effect on accuracy, since the two error curves lie on top of each other.}
\label{f:hf_errors}
\end{figure}

As an additional benchmark, we tested the ASCI approach against alternative truncations. In these truncations, we found the exact ground state of $H(s)$ ahead of time (either at every time step or only at the final time step $s=1$) and identified the determinants whose coefficients in the wavefunction had a magnitude above a certain cutoff. These "good" determinants were then used to build our state space. ASCI was then run on the final Hamiltonian, finding the determinants required to identify its ground state, with ASCI's target size being the number of determinants found above the lowest truncation (also from the final Hamiltonian) for the sake of a fair comparison. We then found the predicted squared overlap of the evolved state with the target ground state at each time step for each truncation, and subtracted these from the predicted overlap in the exact case, for the same amount of orbitals. This truncation comparison was done for LiH, HF, Li\textsubscript{2}, N\textsubscript{2}, and P\textsubscript{2} with  ASCI outperforming them all handily. The truncation errors for HF are shown in \Fig{hf_errors}. 

\subsection{Error From Small Core Size}
\begin{figure}[htb]

\centering
\includegraphics[width=\linewidth]{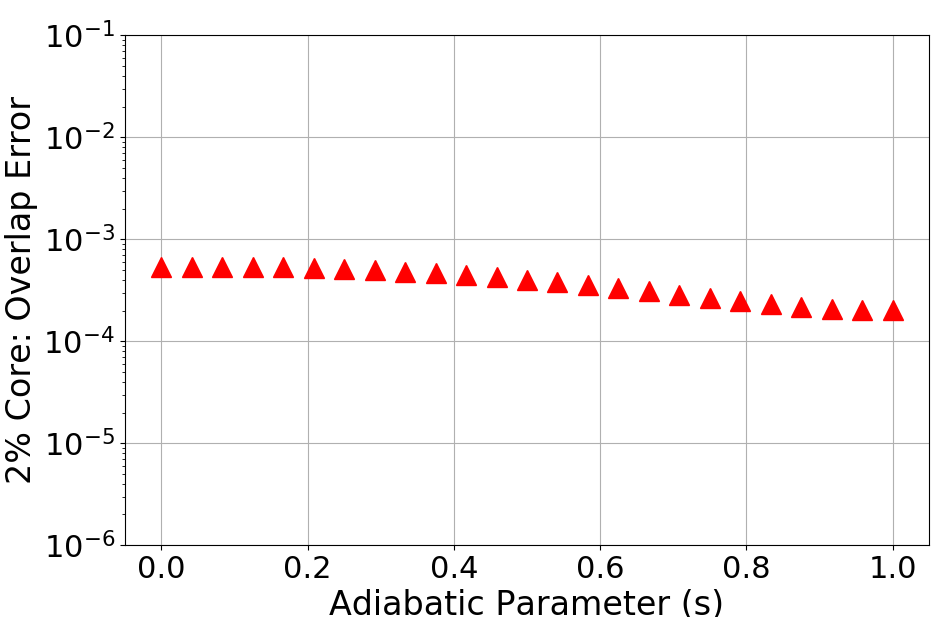}
\caption{Difference in predicted overlap with target ground state for CH\textsubscript{4} between cases where (1) core size = target size and (2) core size = 2\% target size (64 orbitals, 20k target size)}
\label{f:small_core}
\end{figure}

As was mentioned in previous sections, when deciding which determinants to consider for our ASCI basis, we only need to look at determinants formed from single and double excitations of the top few percent of the current iteration's set. This allows us to select a much larger target size and reduce overall error considerably, especially when using the cc-pVTZ orbital basis. However, this does introduce some error on its own, and we tried to estimate it for a few different molecules and target sizes CH\textsubscript{4} (Target Size 20,000), F\textsubscript{2} (Target Size 40,000), P\textsubscript{2}(Target Size 90,000), and CH\textsubscript{2} (Target Size 100,000). The results for CH\textsubscript{4} are showcased in \Fig{small_core}. 
In all cases, the core size was 2$\%$ of the Target Size, and the orbital basis used was cc-pVTZ since this was the only orbital basis for which this approximation was necessary. Overall, the error in initial overlap from this was found to be at most $10^{-3}$, and error in final overlap was found to be at most $2*10^{-4}$.
\subsection{Estimating combined error From ASCI in large full CI spaces}
\subsubsection{Molecular set}
\begin{figure}[htb]

\centering
\includegraphics[width=\linewidth]{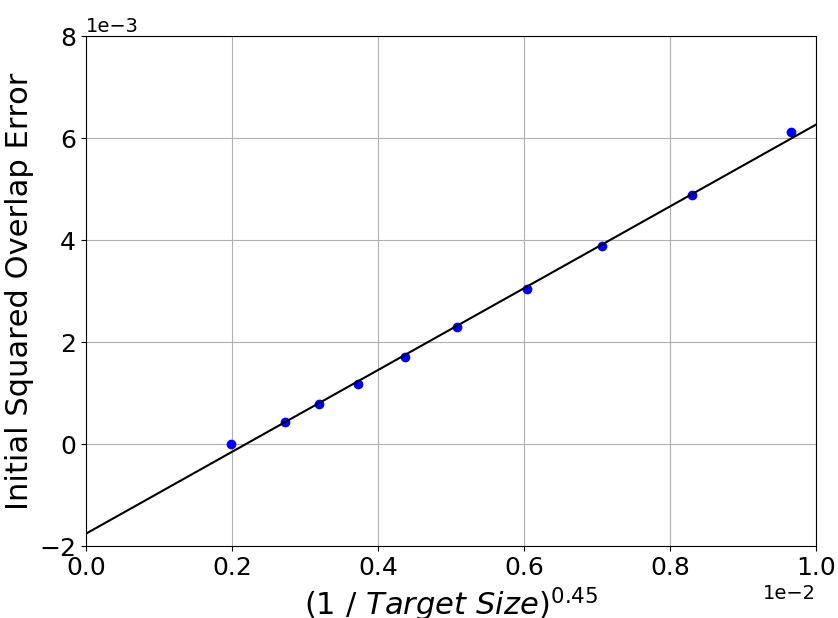}
\caption{Difference in initial squared overlap with ground state between evolutions run with ASCI-generated states (1) using $10^6$ target size and (2) using smaller target sizes. Core size is 2\% of target size. (CH\textsubscript{2}, cc-pVTZ)}
\label{f:initial_asci_error}

\end{figure}

\begin{figure}[htb]

\centering
\includegraphics[width=\linewidth]{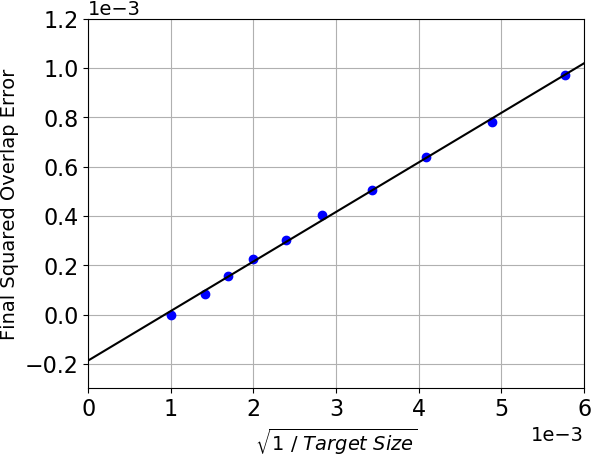}
\caption{Difference in final squared overlap with ground state between evolutions run with ASCI-generated states (1) using $10^6$ target size and (2) using smaller target sizes. Core size is 2\% of target size. (CH\textsubscript{2}, cc-pVTZ)}
\label{f:final_asci_error}

\end{figure}

We also decided to test whether this level of accuracy held for cases where the truncation made up a much smaller portion of the full determinant space. We note that as the target size approaches the size of the full space (that is, as $\frac{1}{\text{target size}} \rightarrow 0$) then we expect the error in overlap to go to zero. Since we cannot compare our predicted overlap to the true overlap , we instead compare our predicted overlaps (for different target sizes) to  the predicted, "reference" overlap using a finite but large number of determinants ($~10^6$). Since the error is positive and monotonically decreasing as target size increases, the approximate error of our data points differs from their true error by the error of our reference overlap (the one calculated using $10^6$ determinants). Thus, the error of all these data points is shifted down by the error of our reference point - so if we could interpolate between the data points and find the y-intercept of the resulting curve, this would reveal the error of our reference points. Since an exact interpolation is infeasible, we tried a simple polynomial interpolation. This approach, tested for small target sizes and cases where we had overlap curves calculated using the full determinant space, was found to give us an error estimate which was 2-20 times larger than the true error. 
 Using this to estimate error in both the initial (ex: \Fig{initial_asci_error}) and final (ex: \Fig{final_asci_error}) overlaps, predicted error upper bounds of $10^{-3}$ for the final overlap and $10^{-2}$ for the initial overlap - more than adequate for our purposes. Thus, using ASCI generated states for time dynamics appears to have a negligible effect on what we aim to find through our simulations. 
 
\subsubsection{Hubbard set}
\begin{figure}[!tbp]
  \centering
  \begin{minipage}[b]{\linewidth}
    \includegraphics[width=\linewidth]{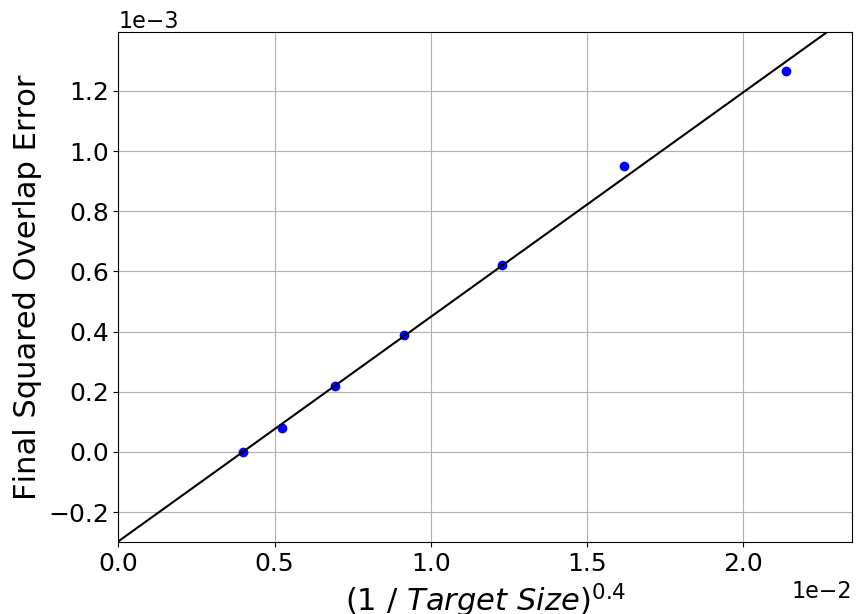}
  \end{minipage}
  \hfill
  \begin{minipage}[b]{\linewidth}
    \includegraphics[width=\linewidth]{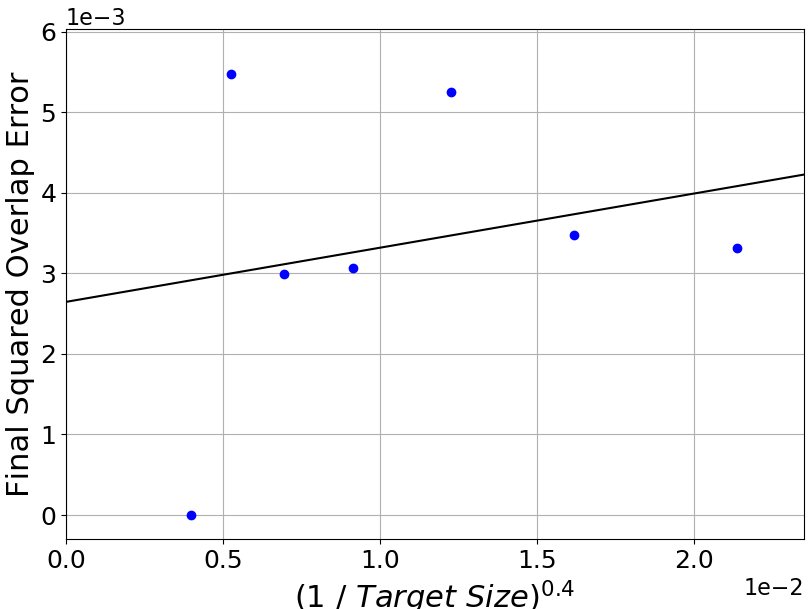}
  \end{minipage}
  \caption{Difference in predicted final overlaps using a target size of $10^6$ vs. using smaller target sizes. Model is a 1D, 18-site lattice at half filling with periodic boundary conditions ($\epsilon = 0$, $t=1$. In the first panel, $U=1$ and there is an initial overlap of $\sim 0.9$. In the second panel $U=3$ and the initial overlap is $\sim 0.45$. }
  \label{f:hubb_ts_different_errs}
\end{figure}

For the Hubbard model, for high values of U, the predicted final overlap with the target ground state no longer monotonically decreases as we increase the size of our truncated space, rendering the interpolation estimate of error undoable. To see this contrast, simply compare the first two graphs of \Fig{hubb_ts_different_errs}.

\subsection{Error in Energy Gap}
\subsubsection{Molecular set}
\begin{figure}[htb]

\centering
\includegraphics[width=\linewidth]{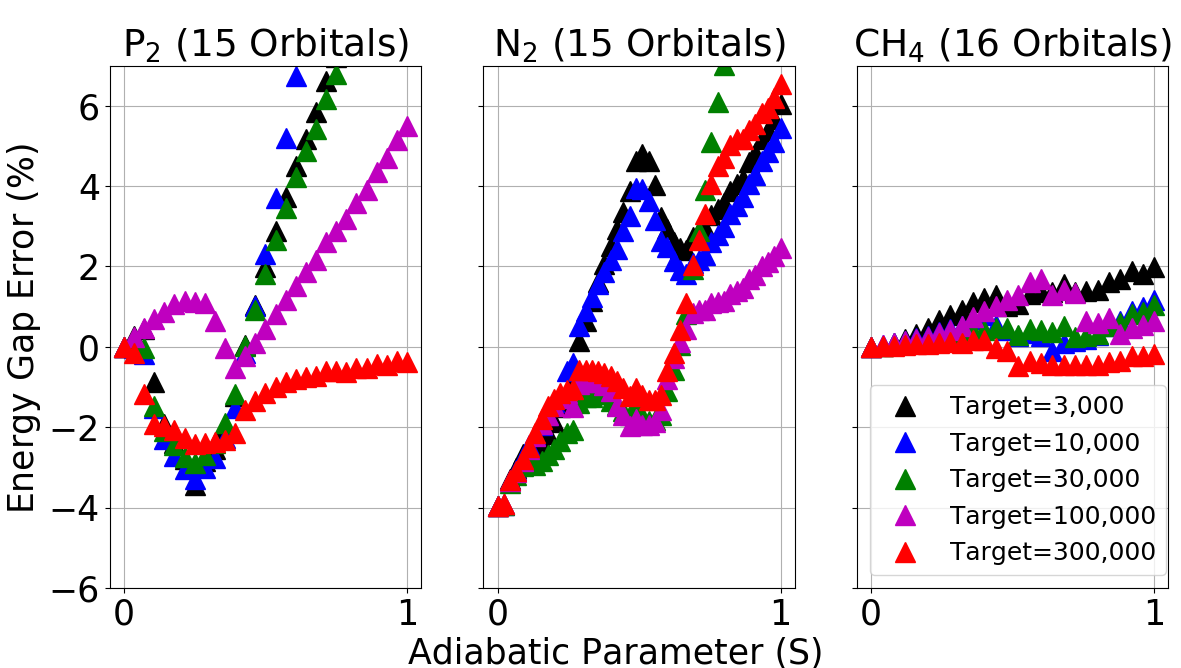}
\caption{Percent error in calculating energy gap using ASCI states. The error at each timestep represents the percent by which the simulation in the truncated determinant space over/under-estimates the true energy gap, as calculated in the full determinant space.}
\label{f:mol_gap_asci_error}

\end{figure}

The percent error in calculating the energy gap  for molecules is displayed in \Fig{mol_gap_asci_error}.

\subsubsection{Hubbard set}
The energy gap error for three Hubbard Models is displayed below. 
\begin{figure}[htb]

\centering
\includegraphics[width=\linewidth]{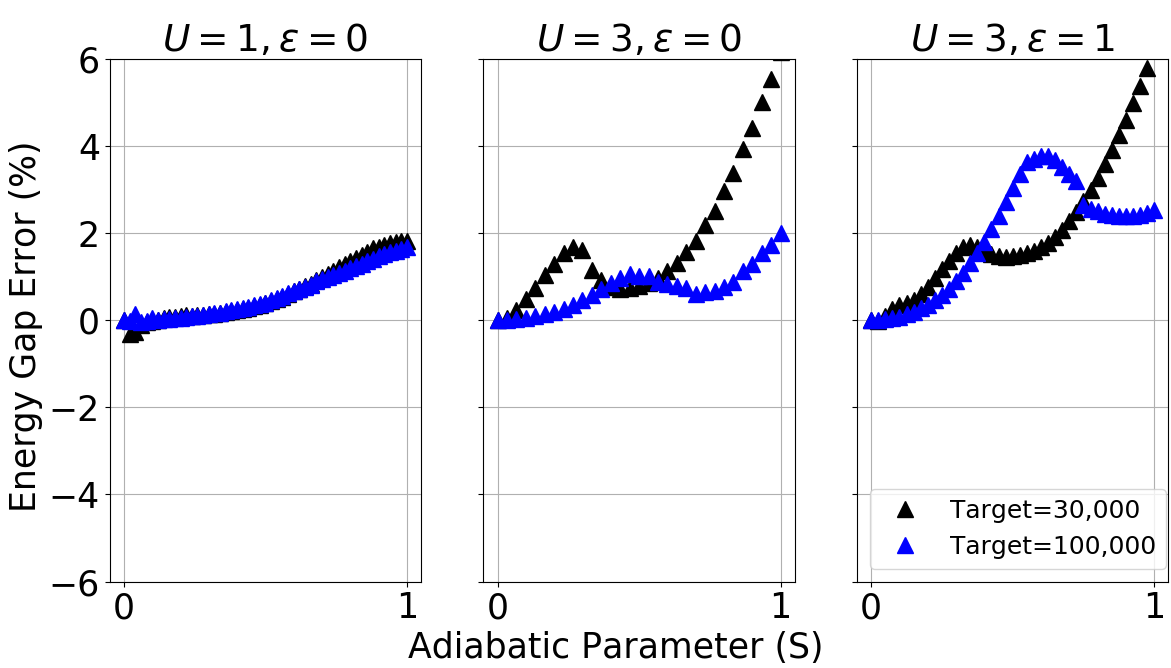}
\caption{Percent error in calculating energy gap using ASCI states. The error at each timestep represents the percent by which the simulation in the truncated determinant space over/under-estimates the true energy gap, as calculated in the full determinant space. The system examined is a 2D, 3x4 lattice at half-filling with periodic boundary conditions, and hopping term $t=1$.}
\label{f:hub_gap_asci_error}

\end{figure}

\end{document}